\documentclass[sigconf,10pt, screen]{acmart}
\renewcommand\footnotetextcopyrightpermission[1]{}  
\AtBeginDocument{%
  }

\usepackage{subcaption}
\usepackage{enumitem}
\begin{document}

\title{MetaBlue: A Metasurface-Assisted Acoustic Underwater Localization System}

\newcommand{\oursystem}{\texttt{MetaBlue} }
\newcommand{\transmitter}{transmitter}
\newcommand{\hydrophone}{hydrophone}
\author{
Junling Wang$^{1}$,
Yi Guo$^{1}$,
Bojun Yang$^{1}$,
Yazhou Yuan$^{3}$,
Zhenlin An$^{2}$\\
{\small
$^{1}$School of Automation and Intelligent Sensing, Shanghai Jiao Tong University, Shanghai, China\\
$^{2}$School of Computing, University of Georgia, Athens, GA, USA\\
$^{3}$School of Electrical Engineering, Yanshan University, Hebei, China\\
}
}

\renewcommand{\shortauthors}{Wang et al.}

\begin{abstract}
Underwater localization is essential for marine exploration and autonomous underwater operations, yet existing radio-frequency and optical approaches are limited by rapid attenuation or limited visibility. Acoustic sensing remains the most practical choice, but conventional acoustic systems typically rely on large arrays or multiple synchronized anchors, resulting in high hardware costs and complex deployment. This paper introduces a novel low-cost passive acoustic metasurface, \oursystem, explicitly designed for underwater localization, which, when attached to an ordinary ultrasonic transmitter, transforms it into a directional ``super-transmitter." The metasurface embeds direction-dependent spectral patterns into the transmitted waveform, enabling accurate angle-of-arrival (AoA) estimation using only a single hydrophone. For ranging, we present a new EM-acoustic mixed time-of-arrival (ToA) method that leverages the acoustic transducer’s inherent low-frequency EM leakage as a timing reference, enabling precise ranging without shared clocks. This allows complete 3D localization with a single low-cost anchor. We evaluate the system across diverse real-world underwater settings, including pools, tanks, and outdoor environments. Experiments show that our design achieves an average AoA error of 8.7$^\circ$ and 3D localization error of 0.37 m at distances over 10 m. Even with a single anchor, the system maintains 0.73 m precision.
\end{abstract}
\maketitle
\vspace{-6mm}
\section{Introduction}
\label{fig:intro}
Underwater localization is essential for both human divers and autonomous underwater robots. Real-time self-position not only provides critical safety feedback and mission assurance for divers~\cite{chen2023underwater}, but also is the basis for underwater robots to autonomously execute tasks such as cleaning~\cite{mao2025utc}, rescue~\cite{shahria2019underwater}, and gas exploration~\cite{hu2022underwater}. A wide range of underwater localization techniques has been explored over the years, yet each faces fundamental limitations. RF-based systems~\cite{park20163d} suffer from severe attenuation in water, limiting their effective range to only a few decimeters and making them impractical for most real-world scenarios. Optical methods, including visible light~\cite{lin2024uwbeacon}, passive visible markers~\cite{zhang2022u}, and laser-based localization~\cite{karras2007localization}, offer high precision in clear conditions but rapidly degrade under turbidity, low-light environments, and suspended particles—conditions commonly found in natural underwater settings. 
\begin{figure}[t!]
    \centering
    \includegraphics[width=0.8\linewidth]{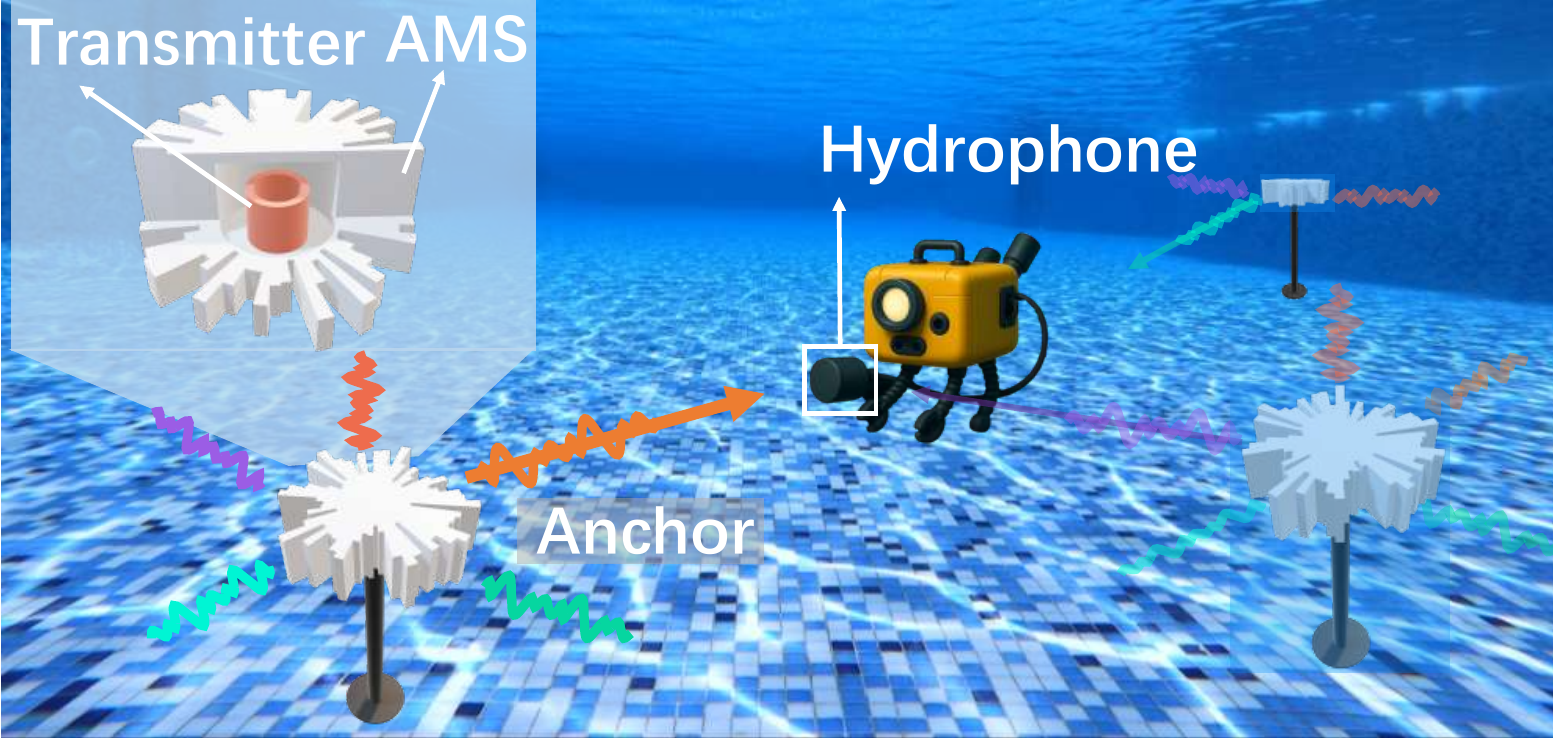}\vspace{-4mm}
    \caption{\oursystem\ Overview. \textnormal{The \textit{passive acoustic metasurface} converts the \transmitter \ into a directional encoder. \textit{A single low-cost hydrophone} can measure direction by decoding the {direction-dependent spectral pattern} (for AoA), and achieve ranging by measuring the {time difference between the electromagnetic leakage and the acoustic signal}. By combining these two measurements, a \textit{single anchor achieves full 3D sub-meter localization}. }}\vspace{-8mm}
    \label{fig:overview}
\end{figure}

\begin{table*}[t!]
\caption{Comparison of SOTA Underwater Localization Systems.}\label{tab:comparison}\vspace{-5mm}\small
\begin{tabular}{l|lllllllll}
\hline
System & Method & Accuracy & Range & Hardware Cost & Bad Visibility & No Clock Sync & One-Anchor\\ \hline
U-Star~\cite{zhang2022u} & Visible light & 1\,m & 3\,m & Passive anchors & $\times$ & \checkmark & \checkmark\\ \hline
UWBeacon~\cite{lin2024uwbeacon} & Visible light & 6\,cm & 10\,m & Multiple LED beacons & $\times$ &\checkmark &$\times$\\ \hline
Chen~\cite{chen2023underwater} & Acoustic ToF & 0.9--1.6\,m & 45\,m & $\geq4$ mic-speaker nodes &  \checkmark & $\times$ & - \\ \hline
3D-Blue~\cite{afzal20243d} & Acoustic AoA & 0.25--1.4\,m & 10\,m & 8 Elements Array & \checkmark & \checkmark & \checkmark \\ \hline
\textbf{Our System} & AoA+ToF & 0.37-0.73\,m & 12\,m & 1 Transducer + AMS & \checkmark & \checkmark & \checkmark \\ \hline
\end{tabular}\vspace{-5mm}
\end{table*}

Acoustic sensing is the dominant solution for underwater applications, broadly categorized into \textit{time-based ranging} and \textit{direction finding}. {Time-based ranging} (ToA~\cite{webster2009preliminary} and TDoA~\cite{ramachandran2023acoustic}) determines position from acoustic travel time, requiring either accurate absolute (ToA) or precise relative (TDoA) clock synchronization across a network of multiple devices. 
This dependency mandates expensive infrastructure and complex deployment.
Chen et al.~\cite{chen2023underwater} provide a TDoA combined with topology-based localization, 
but this still requires frequent message exchanges for synchronization and at least 4 nodes, each node has both a transmitter and 2-3 receivers to achieve reliable positioning.
Conversely, {Direction Finding} (AoA) techniques~\cite{afzal20243d} avoid clock synchronization but necessitate large hydrophone arrays at the receiver, which imposes significant hardware, power, and size constraints on the mobile agent.
For example, 3D-Blue~\cite{afzal20243d} uses an 8-hydrophone array with an aperture of up to 3.71 m for localization.
Thus, existing techniques invariably mandate either complex multi-anchor infrastructure or bulky multi-element receiver hardware, severely limiting scalability for \textit{single-anchor, low-cost deployment}. We summarize these limitations in Table~\ref{tab:comparison}.

Acoustic metasurface (AMS)–based localization systems offer a compelling solution to mitigate the hardware cost and power consumption associated with traditional arrays. Some existing research has established this AMS paradigm in the airborne indoor localization~\cite{wang2025metasonic,garg2021owlet, fu2024pushing}. The core principle is \textit{spatial encoding}: a passive metasurface, integrated either on the transmitter-side (e.g., MetaSonic~\cite{wang2025metasonic}) or the receiver-side (e.g., Owlet~\cite{garg2021owlet}), embeds \textit{direction-dependent spectral signatures} into the sound field, successfully enabling AoA estimation without a bulky multi-element array.

While the acoustic AMS paradigm is well-established for airborne indoor localization, its development for underwater positioning has just started~\cite {li2024metastructure,chen2019broadband,jin2024bubble}. Initial studies, such as those employing complex structural designs like bubble-structured units~\cite{jin2024bubble}, mainly focus on achieving basic beam control. These emerging underwater efforts face three critical limitations. First, they often utilize complex unit cell structures, or necessitate costly materials (e.g., aluminum~\cite{chen2019broadband}) rather than common, 3D-printable plastics. Second, they fundamentally \textit{lack the necessary ranging mechanism} to achieve 3D localization, focusing exclusively on AoA estimation. Third, the resulting systems are typically limited to preliminary single-node verification and have \textit{not} been validated as fully integrated localization systems in real-world environments. Consequently, a clear gap remains for an AMS-based system that is structurally simple, cost-effective, and fully optimized to enable robust 3D localization.

In this paper, we introduce \oursystem, a new metasurface-based underwater localization system that, for the first time, enables full 3D localization using only a single low-cost anchor. Unlike prior AMS-based approaches, \oursystem provides a structurally simple 3D localization solution. As shown in Fig.~\ref{fig:overview}, we encapsulate an ordinary ultrasonic \transmitter within \textit{a passive, low-cost 3D-printed acoustic metasurface} that embeds direction-dependent spectral signatures into the transmitted waveform. A single hydrophone can decode these signatures to estimate the AoA and range. By jointly leveraging these measurements, \oursystem delivers robust underwater 3D localization with minimal hardware. However, translating this idea into a practical  system introduces three key challenges:

The first challenge arises from a fundamental material mismatch. Airborne AMS designs commonly rely on low-cost 3D-printed plastics~\cite{rathod2020review}, which work well in air because the large air–solid impedance contrast enables waveform modulation for phase control~\cite{zhang2023acoustic}. Underwater, however, water’s impedance closely matches that of 3D-printed materials, causing most acoustic energy to pass through the metasurface rather than be modulated. To overcome this, we develop a new metasurface mechanism that controls the exit phase at the unit-cell level through thickness-engineered solid and water layers, and jointly optimizes these thicknesses at the array level to maximize the directional diversity of the radiated sound field. (Sec.~\ref{sec:metasurface})

Second, the severe multipath in shallow-water environments. Reflections from the surface, bottom, and surrounding structures create multiple NLOS paths, causing LOS and NLOS components to overlap and distort the directional spectral features used for AoA, as detailed in Sec .~\ref {subsec:multipath_analysis}. This overlap fundamentally degrades the accuracy of AoA estimation. To address this, we first estimate the minimum LOS–NLOS TDoA and then design a chirp-based multipath-suppression algorithm, parametrized by this lower bound, to mitigate multipath interference.

The third challenge is accurate range estimation without complex synchronization. Classic handshake-based two-way ranging schemes~\cite{peng2007beepbeep} suffer from timing granularity limits, ToA jitter, and protocol delays when using short acoustic probes, making their timing uncertainty comparable to the acoustic propagation time and thus unsuitable for lightweight deployments. We address this by leveraging the acoustic \transmitter’s inherent ultra-low-frequency electromagnetic leakage signal~\cite{gong2024enabling} as an instant timing reference, enabling synchronization-free, accurate one-way ranging.

We design and implement \oursystem and make following contributions: 
\begin{itemize}[leftmargin=10pt]
\item 
We design the first low-cost, parameter-optimized AMS that is effective in water and dedicated to AoA estimation. The result shows that the AMS enables AoA estimation using a single hydrophone, improving the accuracy by 55\% compared to a baseline without AMS. The cost of our AMS is only about \$6.
\item We propose a multipath suppression algorithm and model the spatial distribution of LOS-NLOS TDoA. The algorithm can recover the spectral features even when LOS and NLOS components are heavily overlapped. The result shows our algorithm improves AoA estimation accuracy of 77\%, with a mean AoA error of 8.7 $^\circ$ in a 360$^\circ$ range.

\item We develop a 3D localization framework that jointly leverages EM–acoustic TDoA ranging, AoA, and depth estimations. This framework enables one-anchor, synchronization-free localization. \oursystem achieves a median localization error of 0.73 m with one anchor and  0.37 m with four anchors. The accuracy is comparable to recent works with far less hardware cost, as shown in Table~\ref {tab:comparison}.
\end{itemize}

\begin{figure}[t]
\centering
\begin{subfigure}[t]{0.3\textwidth}
    \centering
    \includegraphics[width=\textwidth]{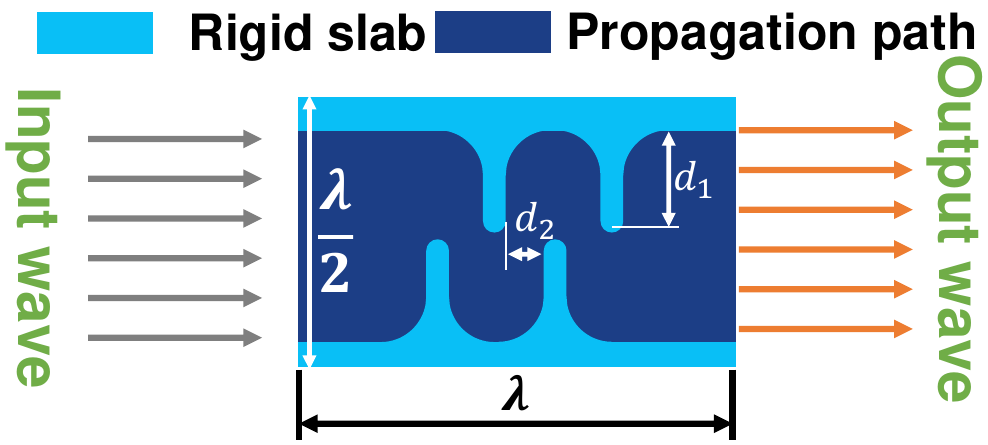}
    \caption{\small AMS unit cell structure.} 
    \label{fig:traditional_ams_unit}
\end{subfigure}
\begin{subfigure}[t]{0.24\textwidth}
    \centering
    \includegraphics[width=\textwidth]{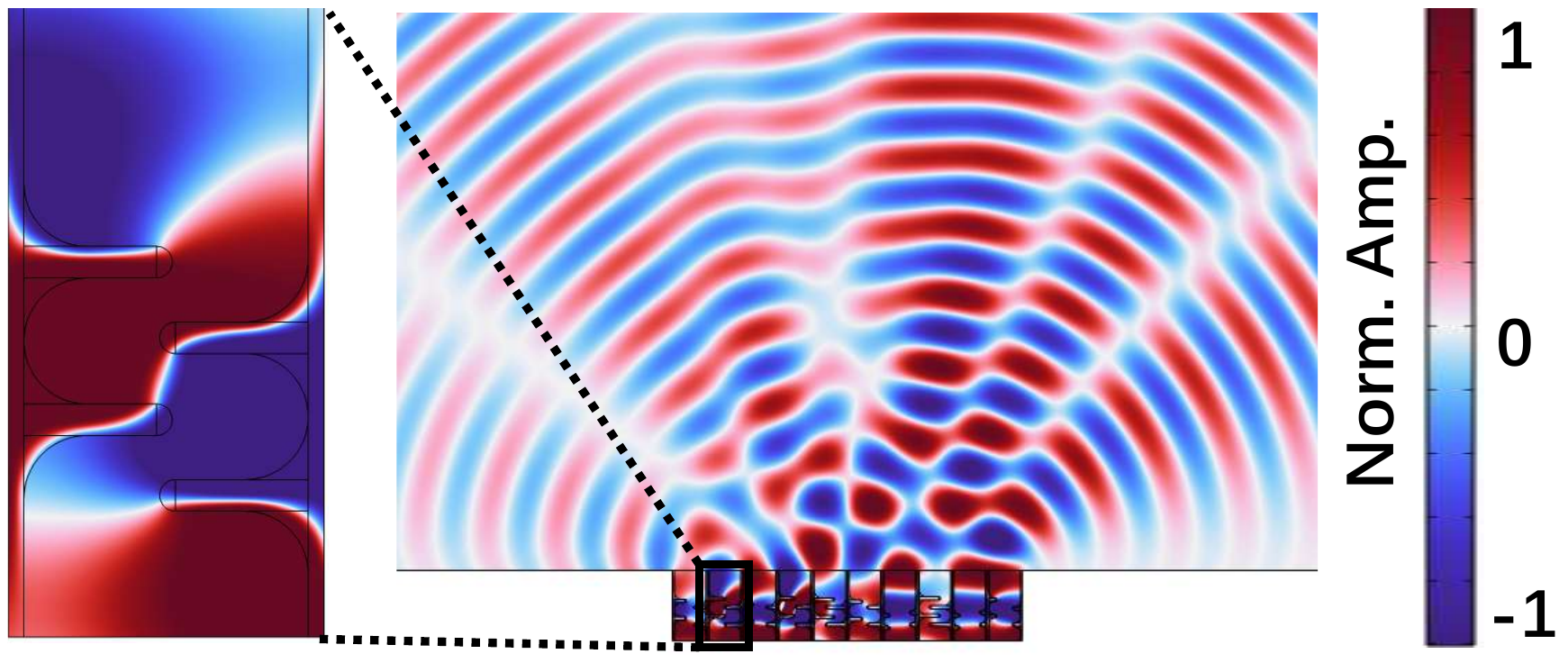}
    \caption{Air}
    \label{fig:air_ams_in_air}
\end{subfigure}%
\begin{subfigure}[t]{0.24\textwidth}
    \centering
    \includegraphics[width=\textwidth]{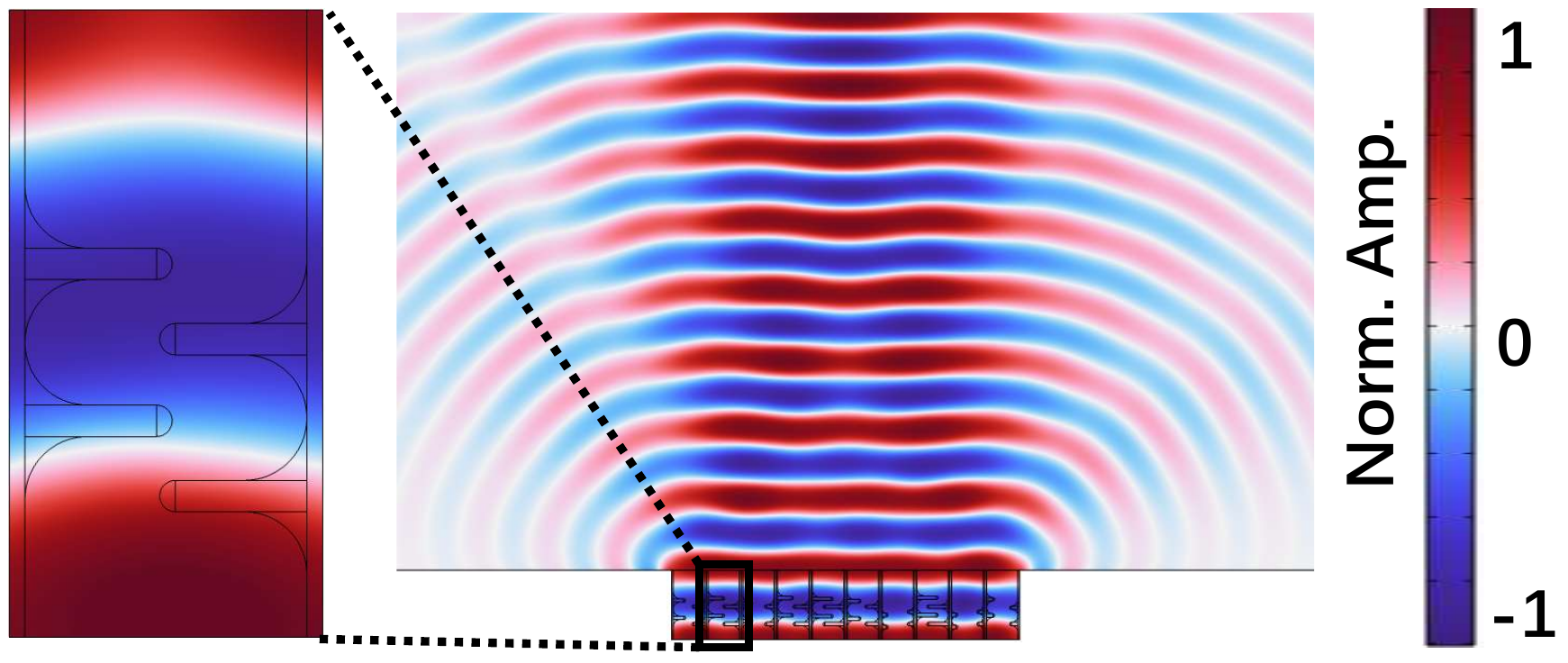}
    \caption{Water}
    \label{fig:air_ams_in_water}
\end{subfigure}\vspace{-5mm}
\caption{Past AMS. \textnormal{(a) The structure of a unit cell. The $\lambda$ is the wavelength and the phase shift is determined by $d_1$ and $d_2$.
(b)–(c) Instantaneous sound pressure fields of a single AMS unit cell and the AMS array in air and water.}}
\label{fig:air_unit_simulation}\vspace{-7mm}
\end{figure}
\section{Preliminary}
\subsection{AMS for AoA Localization Systems}
Recently, AMS, which shapes the sound field of acoustic devices, has been extensively explored as a low-cost solution for AoA localization. An AMS~\cite{assouar2018acoustic} consists of subwavelength unit cells that enable precise phase or amplitude control. A widely used AMS unit cell is shown in Fig.~\ref{fig:traditional_ams_unit}. Its inner channel contains two parametrically designed curled pillars, controlled by $d_1$ and $d_2$, which shape the acoustic propagation path. As the incident wave traverses the cell, these pillars extend the effective path length relative to a pillar-free channel, thereby introducing a controllable phase shift on the transmitted wave. By varying $d_1$ and $d_2$, each unit cell can be assigned a distinct phase shift at the design frequency. 
When the unit cells are arranged into an array to form the AMS, each cell can be regarded as a secondary acoustic source with its own phase and amplitude. Together, these secondary sources form an array whose coherent far-field superposition yields a direction-dependent sound-pressure distribution.

Here, we use COMSOL~\cite{comsol2024} simulation to show the direction-dependent sound pressure. 
First, we construct a 10-cell PLA-made AMS\footnote{PLA stands for Polylactic Acid, a low-cost and widely used 3D-printing material.}, with $\lambda = 2 \text{cm}$ and 10 parameter pairs of random $(d_1,d_2)$.
Then, we set the AMS in the air and subject it to an incident plane wave with a wavelength of 2~cm.
Fig.~\ref{fig:air_ams_in_air} shows the simulated instantaneous sound pressure field. The color encodes instantaneous acoustic pressure, with darker shades indicating higher magnitude. Red and blue represent positive and negative pressure, respectively, and the propagation direction is perpendicular to the isobars, following the alternating red–blue pattern.
The left panel shows a zoomed-in view of a single unit cell. The sound wave is guided along the channel constrained by the two pillars. We measure a total phase shift of $\tfrac{11}{3}\pi$ between the input and output waves. After subtracting the baseline $2\pi$ phase shift from propagation over one wavelength, the internal structure contributes an additional $\tfrac{5}{3}\pi$ phase shift.
The right panel shows the sound field of the phase-engineered 10-cell AMS array. The variation in harmonic-wave amplitude with angle reveals a direction-dependent sound pressure.
The above analysis shows how the AMS imprints directional features onto the sound field.
\begin{figure*}[t!]
    \centering
\begin{subfigure}[t]{0.27\textwidth}
    \centering
    \includegraphics[width=\textwidth]{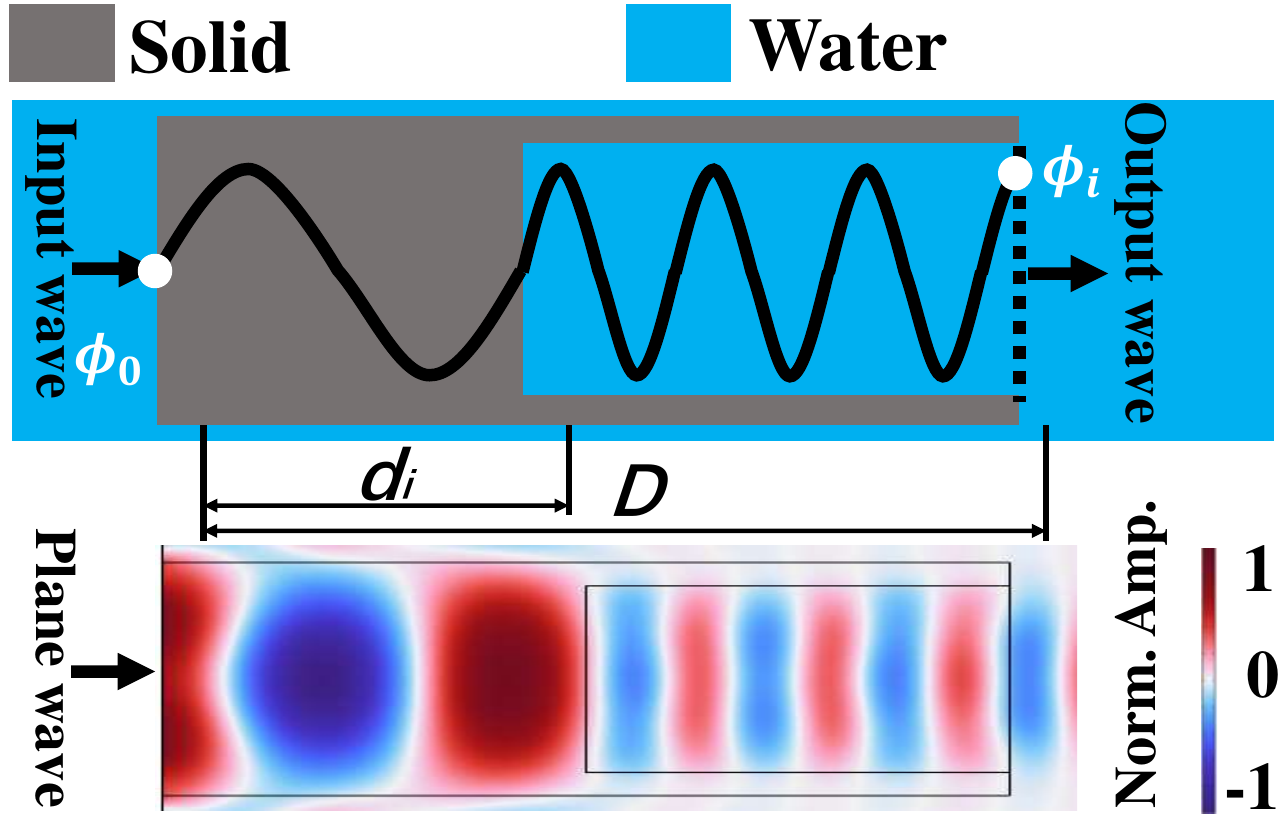}
    \caption{Waterborne AMS unit cell}
    \label{fig:unit}
\end{subfigure}
\begin{subfigure}[t]{0.225\textwidth}
    \centering
    \includegraphics[width=\textwidth]{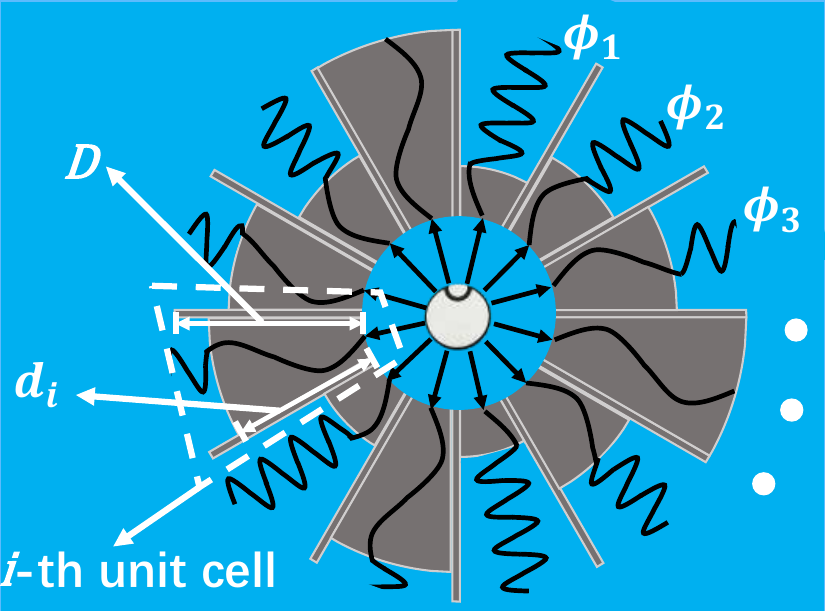}
    \caption{AMS model}
    \label{fig:waterborne_model}
\end{subfigure}
\begin{subfigure}[t]{0.22\textwidth}
    \centering
    \includegraphics[width=\textwidth]{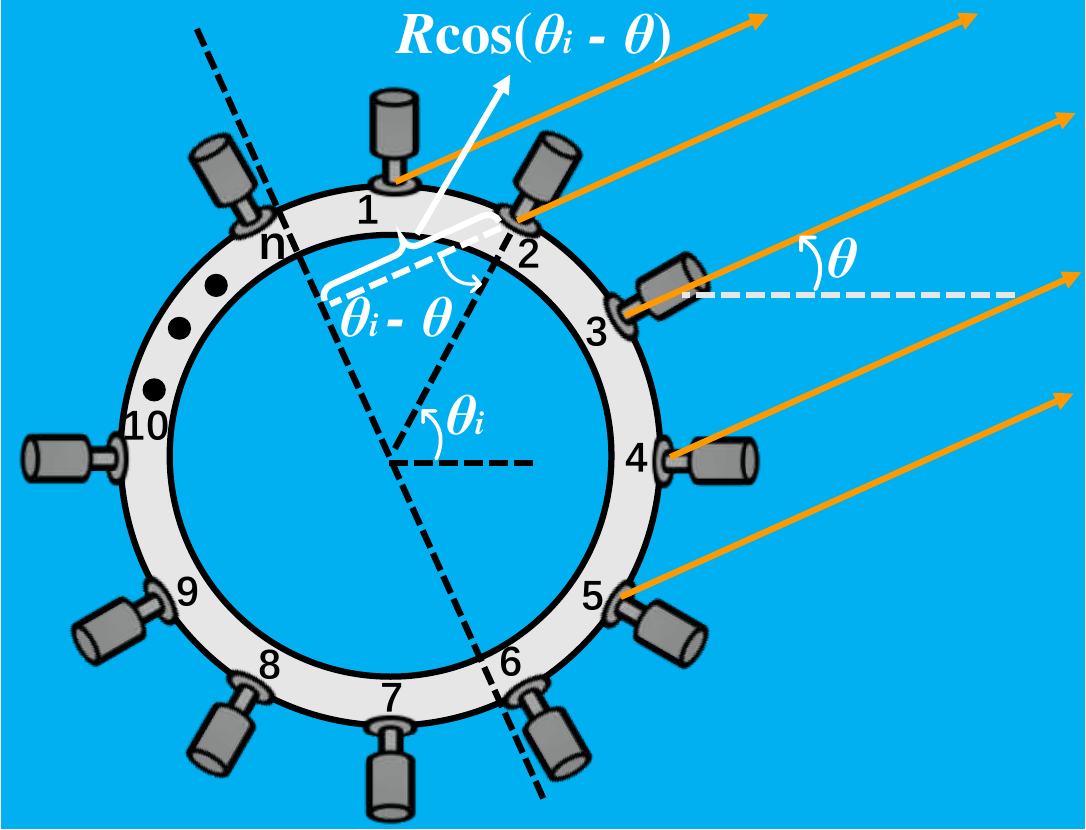}
    \caption{Far-filed wave}
    \label{fig:far_field_wave}
\end{subfigure}
\begin{subfigure}[t]{0.255\textwidth}
    \centering
    \includegraphics[width=\textwidth]{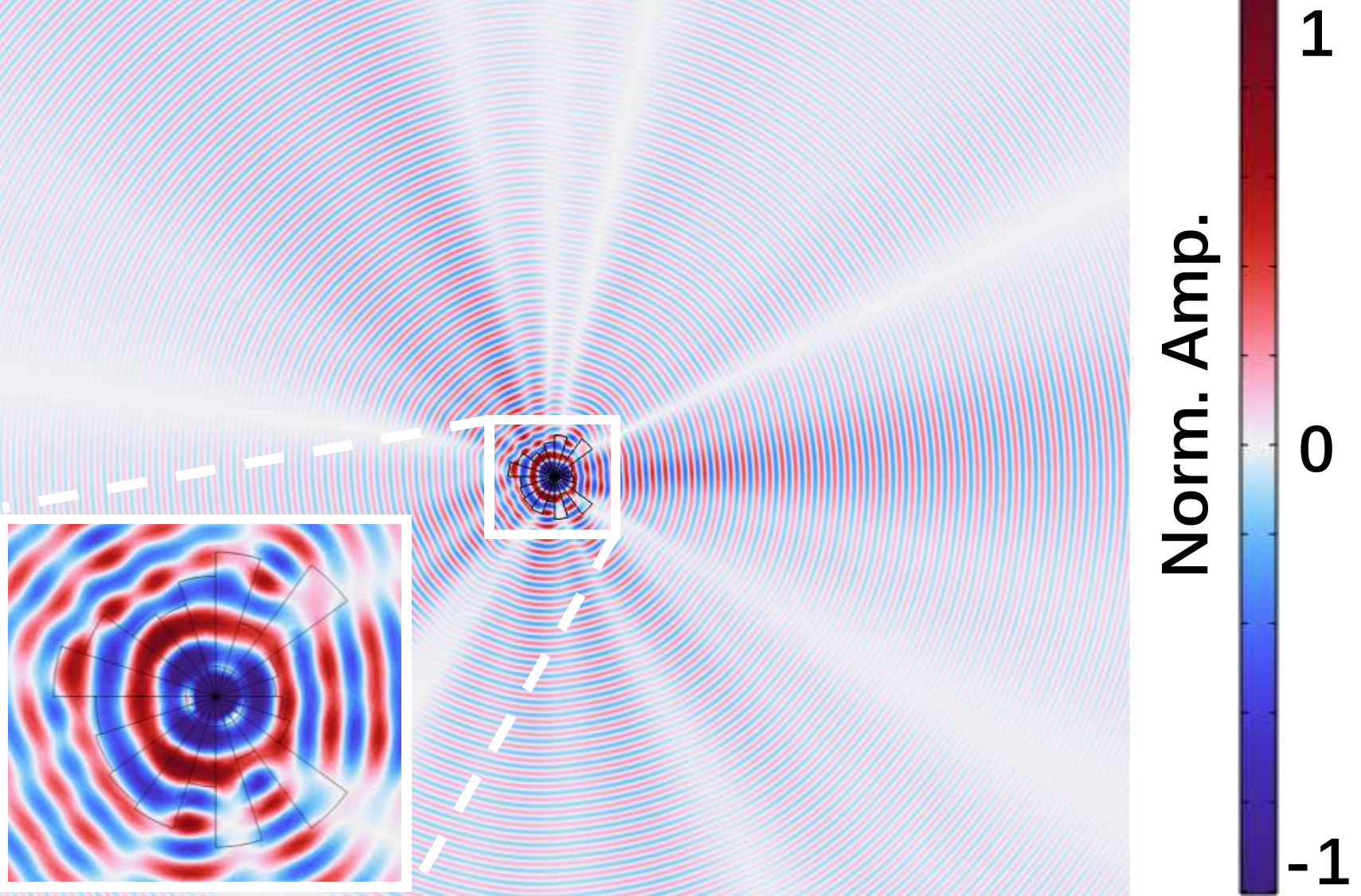}
    \caption{Directional diversity}
    \label{fig:far_field_simulation}
\end{subfigure}
\label{fig:water_AMS_unit}\vspace{-5mm}
\caption{Waterborne AMS Design. \textnormal{(a) The top shows the structure of the waterborne unit cell, and the bottom shows instantaneous sound pressure with an input of a plane wave. (b) shows the circular AMS array. (c) shows the equivalent far-field wave of the circular array. (d) shows the simulated directional sound pressure of a 20-cell AMS driven by a 196 kHz source.}}\vspace{-5mm}
\end{figure*}
\subsection{Challenges of Underwater AMS}\label{sec:challenge}
Although the AMS performs well in airborne environments, it becomes ineffective in waterborne environments.
The simulation is conduct in water, as shown in Fig.~\ref{fig:air_ams_in_water}. 
We reuse the same AMS configuration as in the air case and the incident plane wave has a 2 cm wavelength in water.
The zoomed-in view in the left panel shows that the wave passes through the unit cell nearly uniformly, with negligible influence from the internal structure. Phase measurements at the cell output indicate a total phase shift of approximately $2\pi$, corresponding to propagation over one wavelength length; in other words, the inner structure contributes essentially no additional phase shift. Consequently, in the right panel, the far-field sound pressure pattern closely resembles the natural radiation of the bare plane wave: the wavefront remains approximately planar in front of the unit array and exhibits nearly no directional features.

The ineffectiveness of traditional AMS designs in water stems from the acoustic impedance match between water and solid materials: the acoustic impedance of water is about 3600 times that of air and is close to that of common 3D-printed materials~\cite{antoniou2021acoustical}. As a result, a water–solid interface no longer behaves like the rigid, highly reflective boundary seen at an air–solid interface. Instead, it allows most of the acoustic energy to transmit into the solid.
The power transmission and reflection at the interface between two media~\cite{brekhovskikh2012acoustics} are given by
\begin{equation}\small
    R_E = \frac{(Z_2 - Z_1)^2}{(Z_2 + Z_1)^2}, \quad
    T_E = \frac{4 Z_1 Z_2}{(Z_2 + Z_1)^2},
    \label{eq:transmision_portion}
\end{equation}
where $Z_1$ and $Z_2$ are the acoustic impedances of the two media. Here, $R_E$ and $T_E$ denote the fractions of incident acoustic power that are reflected and transmitted, respectively.
For a water–PLA interface with water's impedance $Z_1 \approx 1.45~\text{MRayl}$~\cite{antoniou2021acoustical}. PLA's impedance $Z_2 = 2.285~\text{MRayl}$~\cite{antoniou2021acoustical}, Eq.~\eqref{eq:transmision_portion} yields approximately $R_E \approx 5\%$ and $T_E \approx 95\%$, i.e., only a small portion of the energy is reflected while the vast majority is transmitted across the boundary. In contrast, for an air–PLA interface with $Z_1 \approx 0.0004~\text{MRayl}$~\cite{rathod2020review} and the same $Z_2$, the reflection and transmission coefficients are about $R_E \approx 99.93\%$ and $T_E \approx 0.07\%$, indicating an almost perfectly reflecting boundary. This stark difference explains why AMS designs that rely on strong air–solid impedance mismatch fail to provide effective phase control when simply immersed in water.

\begin{figure}[t!]
    \centering
    \includegraphics[width=0.8\linewidth]{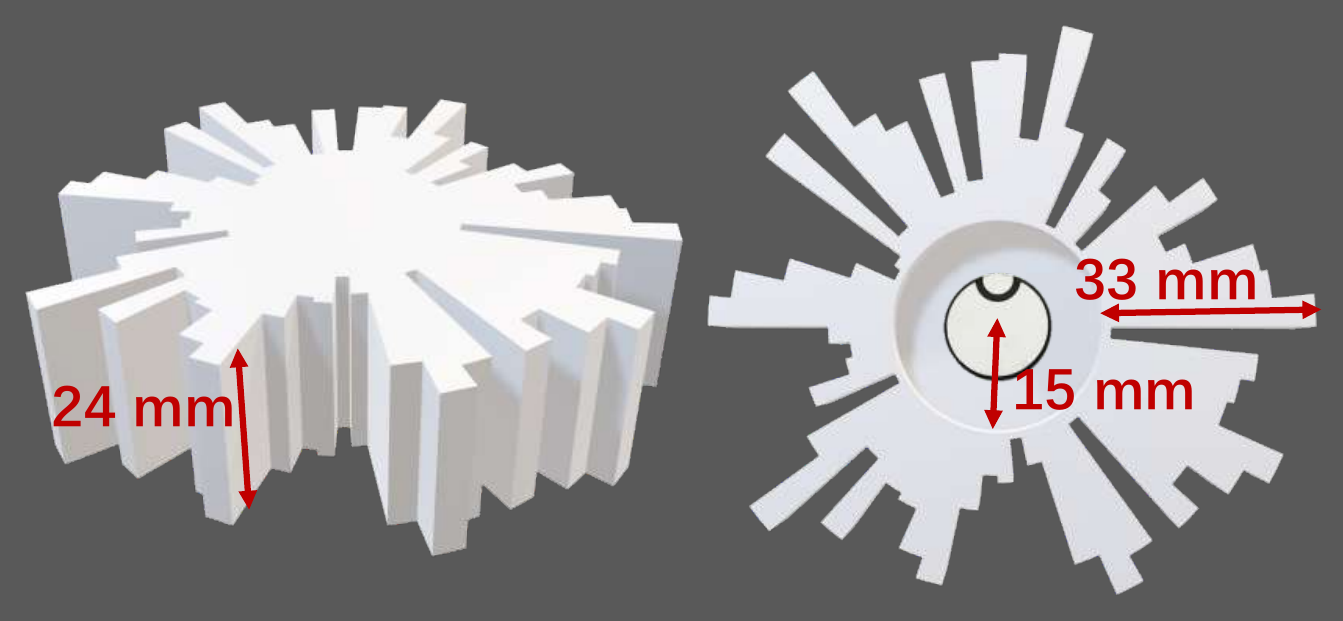}\vspace{-3mm}
    \caption{3D Model of AMS Structure.}\vspace{-5mm}
    \label{fig:3d_AMS_structure}
\end{figure}
\section{System Design}
We present the design of \oursystem, beginning with our waterborne AMS and the model for its directional features. Next, we introduce a multipath suppression algorithm for accurate AoA estimation and our EM–acoustic ranging scheme. The section concludes with a description of the scalable 3D localization method that fuses these range and AoA estimates.
\subsection{Acoustic Metasurface Design}\label{sec:metasurface}
In this part, we detail the design of our underwater AMS. We first describe the unit cell design principle, then derive the far-field sound pressure of the unit-cell array, and finally introduce an optimization method to obtain an AMS configuration that maximizes directional features.

\noindent $\blacksquare$ \textbf{Underwater AMS Unit Cell}.
To realize an effective AMS, we first need to design a unit cell capable of modulating the phase of the incident sound wave. Our design is inspired by the underlying acoustic properties of heterogeneous media: different media exhibit different acoustic wavelengths, so by stacking two materials and tuning their layer thicknesses—and thus the propagation path lengths in each medium—we can impose the desired phase shift.
As shown is Fig.~\ref{fig:unit},
The unit cell is a bar-shaped structure composed of two parts: a solid segment of length $d_i$ and a water-filled segment of length $ D-d_i$, where $D$ is the fixed total length. In the two parts, the acoustic wavelengths differ due to their different acoustic properties. By adjusting $d_i$, the output phase of the transmitted wave can be precisely controlled, as given by:
\begin{equation}\small
\begin{aligned}
\varphi_i(f,d_i) &= \left[ \left( \frac{d_i}{\lambda_1(f)} + \frac{D - d_i}{\lambda_2(f)} \right) \bmod 1 \right] \times 2\pi,
\end{aligned}
\label{eq:unit_phase_delay}
\end{equation}
where \(\varphi_i(f,d_i)\) denotes the phase shift, \(\lambda_1(f) = \frac{c_1}{f}\) is the wavelength in the solid material, and \(\lambda_2(f) = \frac{c_2}{f}\) is the wavelength in water. Here, \(c_1\) and \(c_2\) represent the sound speeds in the solid and water, respectively. The phase $\varphi_i(f,d_i)$ varies within the range \(\left[\frac{2\pi D f}{c_1}, \frac{2\pi D f}{c_2}\right]\). To achieve a full phase coverage of \([0, 2\pi]\), the phase variation must satisfy 
\begin{equation}
    \frac{2\pi D f}{c_2} - \frac{2\pi D f}{c_1} > 2\pi
\end{equation}
which leads to the condition
\begin{equation}
    D > \frac{c_1 c_2}{(c_2 - c_1) f}
    \label{eq:unit_cell_size_condition}
\end{equation}

In addition to the phase shift, acoustic propagation through the unit cell also induces amplitude attenuation, mainly due to absorption in the solid material. Following the frequency-dependent absorption model in~\cite{antoniou2021acoustical}, the attenuation coefficient $\alpha(f)$ (in dB/cm) is given by
\begin{equation}
    \alpha(f) = C f^{n},
    \label{eq:db_attenuation}
\end{equation}
where $C$ is the attenuation prefactor and $n$ is the material-dependent power-law exponent. For a slab of thickness $d$ at frequency $f$, the corresponding amplitude transmission factor, i.e., the ratio between the transmitted and incident pressure amplitudes, is
\begin{equation}
    A(f,d) = 10^{-\frac{d\,C f^{n}}{20}}.
    \label{eq:amplitude_transmission}
\end{equation}
Based on the above principles, we use a PLA–water stack as a representative unit cell to modulate a sound wave of frequency $f = 200~\text{kHz}$. First, using Eq.~\ref{eq:unit_cell_size_condition}, we determine the unit-cell thickness $D$. PLA has a sound speed of $c_1 = 1939.4~\mathrm{m/s}$~\cite{antoniou2021acoustical}, and water $c_2 = 1500~\mathrm{m/s}$~\cite{noaa_underwater_sound}. For a 200~kHz wave, a unit-cell size of $D = 3.3~\mathrm{cm}$ is sufficient to realize a full $0$–$2\pi$ phase modulation.
For amplitude attenuation, following the experimental PLA model in~\cite{antoniou2021acoustical}, we use $n = 1.39$ and
$C \approx 3.72 \times 10^{-8}~\mathrm{dB}\,\mathrm{cm}^{-1}\,\mathrm{Hz}^{-n}$ in Eq.~\ref{eq:amplitude_transmission}, which yields an attenuation coefficient of $0.87~\mathrm{dB/cm}$ at $200~\mathrm{kHz}$. Consequently, for PLA thicknesses $d_i$ of 1~cm, 2~cm, and 3.3~cm, the phase shifts are
$\varphi_i = 0.62~\text{rad}$, $5.00~\text{rad}$, and $2.53~\text{rad}$, respectively, showing that tuning $d_i$ effectively controls the output phase. The corresponding amplitude transmission factors are 90.5\%, 81.9\%, and 71.9\%, indicating that most of the signal amplitude is preserved and the unit cell maintains high transmission efficiency.

\noindent $\blacksquare$ \textbf{Underwater AMS Array}.
After designing the AMS unit cell, we assemble the cells into a circular array to encode direction-dependent features across the full $360^\circ$ azimuth.
As shown in Fig.~\ref{fig:waterborne_model}, The AMS is a set of annular sectors. Each unit cell corresponds to a sector with a fixed inner radius and a variable outer radius, yielding an adjustable solid path length $d_i$ ($i=1,2,\ldots,N$) along the radial direction, where $N$ is the total number of units. 
A PZT \transmitter \ is placed at the center of the AMS to generate an ultrasound signal, which is modulated by the structure and then radiated into the external water medium. 
The actual 3D AMS model, with an inner radius of 15~mm, maximum outer radius of 48~mm, and a height of 24~mm, is shown in Fig.~\ref{fig:3d_AMS_structure}.

To construct directional features, we assign different lengths $d_i$ to different unit cells. However, arbitrary lengths do not necessarily yield an AMS with high directional diversity, which is crucial for accurate direction estimation. To determine an optimal configuration, we first develop an analytical model of the AMS far-field sound pressure.

As illustrated in Fig.~\ref{fig:far_field_wave}, we model each unit cell as an independent secondary source with its own phase and amplitude. We consider a point source at frequency $f$ located at the array center, which generates an incoming field with complex pressure $P_0(f)$ at the entrance of each unit cell.
The outgoing sound pressure from the $i$-th unit cell is then
\begin{equation}\small
    P_i(f) = P_0(f)\,A(f,d_i)\,e^{j\varphi_i(f,d_i)}
\end{equation}
where $\varphi_i(f,d_i)$ and $A(f,d_i)$ denote the phase delay and amplitude transmission factor of the $i$-th unit cell, respectively, as defined in Eqs.~\ref{eq:unit_phase_delay} and~\ref{eq:amplitude_transmission}.

By superposing the contributions of the $N/2$ active units, each located at angle $\theta_i$, the far-field sound pressure $p(\theta,f)$ at observation angle $\theta$ is given by
\begin{equation}\footnotesize
\begin{aligned}
  &p(\theta,f) 
  = \sum_{i=p}^{p+N/2} 
     P_i(f)\,
     \exp\!\left(j\,\frac{2\pi r}{\lambda}\cos(\theta-\theta_i)\right)
     \cos(\theta-\theta_i) \\
  &= \sum_{i=p}^{p+N/2} 
     P_0(f)\,A(f,d_i)\,e^{j\varphi_i(f,d_i)}\,
     \exp\!\left(j\,\frac{2\pi r}{\lambda}\cos(\theta-\theta_i)\right)
     \cos(\theta-\theta_i)
\end{aligned}
\label{eq:far_field}
\end{equation}
where $r$ is the maximum outer radius of the AMS, and $\theta_i$ denotes the angular position of the $i$-th unit. The exponential term 
$\exp\!\left(j\,\frac{2\pi r}{\lambda}\cos(\theta-\theta_i)\right)$ accounts for the far-field phase difference between the $i$-th unit and the array center.
As shown in Fig.~\ref{fig:far_field_simulation}, when a point source emits a monochromatic wave, the AMS produces far-field sound pressures that vary significantly with angle, demonstrating substantial directional diversity.

\noindent $\blacksquare$ \textbf{Metasurface Optimization}.
The above analysis shows that our AMS can generate strong directional diversity in the far-field sound pressure and is well modeled by the theoretical eq.~\ref{eq:far_field}. When a multi-frequency (broadband) signal is used, each frequency component experiences the metasurface differently and forms its own direction-dependent pattern. By superposing these components, we obtain a distinctive amplitude spectrum for each direction. 

A numerical simulation is conducted to show the spectral features. We model an AMS comprising $N = 60$ unit cells, with optimized parameters ${d_i}$ selected from the range $[0, 3.3\,\text{cm}]$. A source is placed at the center of the AMS and emits a $0.5\,\text{ms}$ linear chirp as Eq.~\ref{eq:chirp} spanning $100$–$200~\text{kHz}$. The emitted chirp is then shaped by the direction-dependent frequency response obtained from the COMSOL model of the 60-cell AMS, yielding angle-specific chirp spectra.
As shown in Fig.~\ref{fig:spectra_simulation}, the spectra at $0^\circ$, $60^\circ$, and $120^\circ$ exhibit clearly distinct patterns. An AoA estimator can infer the incident angle by analyzing these spectral patterns, and its accuracy improves with the directional diversity. 

This diversity is governed by the set of thickness parameters $\{d_i\}_{i=1}^N$ across the $N$ unit cells. To obtain an AMS configuration with maximal directional diversity, we therefore formulate an optimization problem over $\{d_i\}_{i=1}^N$.
To maximize directional diversity, we seek to minimize the pairwise similarity between distinct directions. We therefore minimize the sum of pairwise similarities $\sum_{i \neq j} G_{i,j}$. The optimization problem is
\begin{equation}\small
\begin{aligned}
\min_{d_1,\ldots,d_N}\quad & \sum_{i \neq j} G_{i,j} + \beta G_{\text{max}} \\
\text{s.t.}\quad & 0 \le d_i \le D,\quad i = 1,2,\ldots,N,
\end{aligned}
\label{eq:optimization}
\end{equation}
where $\beta$ is a weighting factor that trades off the global similarity and the worst-case term $G_{\max} = \max_{i,j} G_{i,j}$. By solving this optimization, we obtain the optimal parameters of the unit cells; a detailed example is provided in Appendix~\ref{app:optimization}.

\subsection{Underwater Direction Finding}\label{sec:multipath}
After obtaining an effective AMS with optimized directional spectral features, we next design an AoA estimation method that exploits these directional cues. In this section, we first describe AoA estimation from the raw spectrum in the absence of multipath. Then, we analyze the extent of multipath distortion and finally introduce a multipath suppression algorithm for robust AoA estimation.

\noindent $\blacksquare$ \textbf{Spatial Spectrum for AoA Estimation.} 
\begin{figure}[t]
    \centering
\begin{subfigure}[t]{0.25\textwidth}
    \centering
    \includegraphics[width=\textwidth]{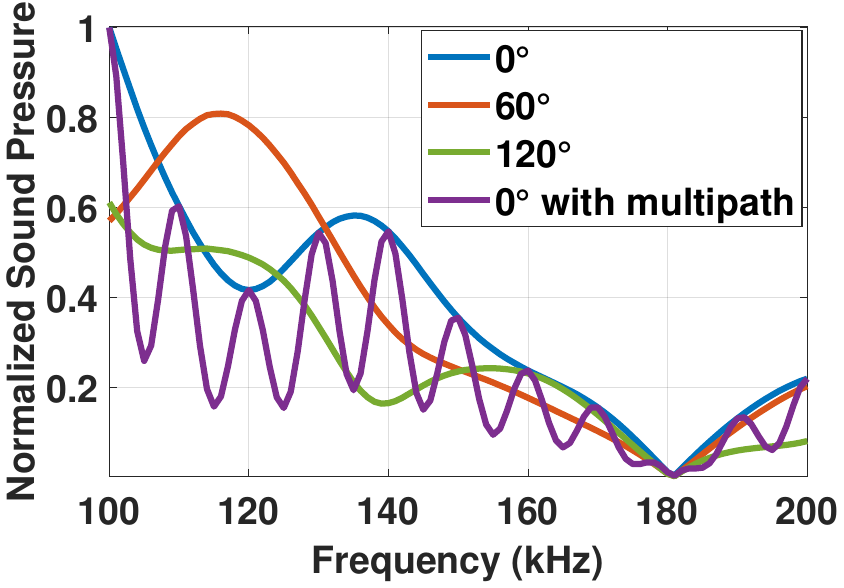}\vspace{-2mm}
    \caption{Spectrum feature}
    \label{fig:spectra_simulation}
\end{subfigure}%
\begin{subfigure}[t]{0.25\textwidth}
    \centering
    \includegraphics[width=\textwidth]{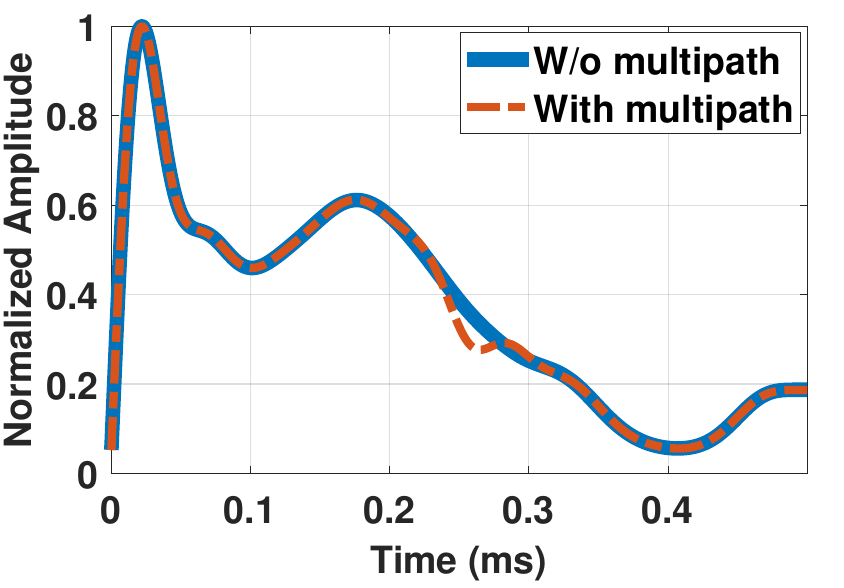}\vspace{-2mm}
    \caption{Filtered signal}
    \label{fig:lowpass_filterd_time_multipath}
\end{subfigure}\vspace{-5mm}
\caption{Directional Feature.\textnormal{
(a) shows raw spectral features at different directions, including cases with and without multipath.
(b) shows signal waveforms with and without multipath after applying the proposed suppression.}}\vspace{-5mm}
\end{figure}
In this section, we describe how the anchor embeds directional features into a broadband signal and how the receiver subsequently probes the signal to estimate the AoA.

For feature embedding, we use a chirp signal because its strong autocorrelation facilitates probing.
The original chirp is emitted uniformly from the transmitter in all directions, and can be expressed as:
\begin{equation}
s(t) = A \cos\!\left( 2\pi\Big((f_0 + \tfrac{1}{2} k t)t\Big)\right), \quad 0 \le t \le T.
\label{eq:chirp}
\end{equation}
where $f_0$ is the starting frequency, $k$ is the slope and $T$ is the duration.
After emission, the chirp propagates through the AMS in all directions. For a given angle $\theta$, it is shaped by the direction-dependent gain $G_\theta(f)$. The resulting output spectrum $S_\theta(f)$ at angle $\theta$ is given by
\begin{equation}
S_\theta(f) = G_\theta(2\pi f) S(2\pi f),
\label{eq:directional_signal_frequency}
\end{equation}
where $S(f)$ is the spectrum of the original chirp.

At the receiver, we first probe the chirp in the received signal by correlating it with the known chirp $s(t)$ and locating the correlation peak~\cite{wang2025metasonic}.
Then, we extract the shaped chirp $r(t)$ from the received signal and estimate the AoA by analyzing its spectrum $R(f)$. The AoA is obtained by matching $R(f)$ against a pre-calibrated template library $\{S_k(f)\}_{k=1}^M$:
\begin{equation}
 \hat{\theta}
= \theta^{(\hat{k})}, \quad
\hat{k}
= \arg\max_{k} \operatorname{Sim}\bigl(R(f), S_k(f)\bigr).
\end{equation}
where $S_k(f)$ denotes the template spectrum corresponding to the $k$-th emitted angle $\theta^{(k)}$ relative to anchor, and $\operatorname{Sim}(\cdot,\cdot)$ denotes a spectral similarity measure.

\begin{figure*}[t]
\centering
\begin{subfigure}[t]{0.25\textwidth}
    \centering    \includegraphics[width=\textwidth]{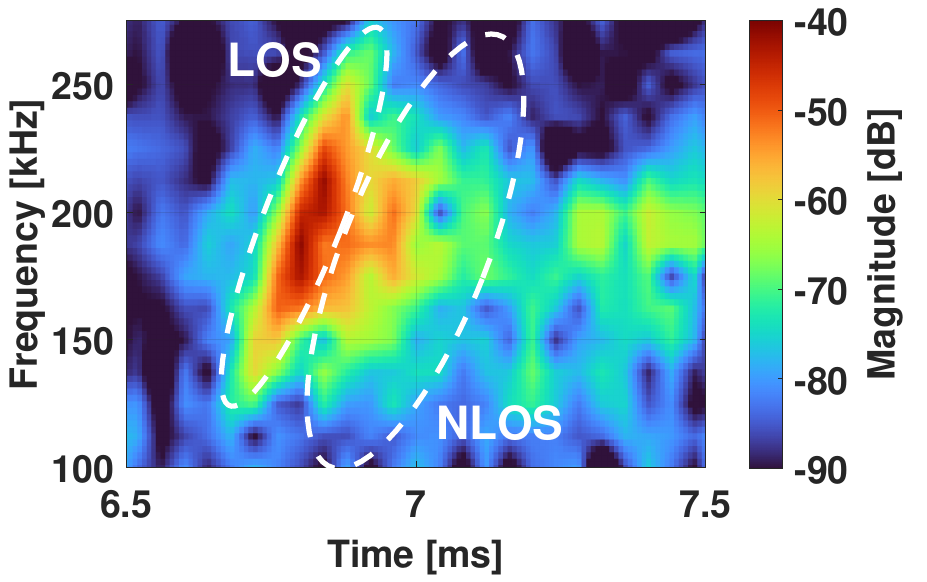}
    \caption{Received chirp}
\label{fig:real_world_raw_signal}
\end{subfigure}%
\begin{subfigure}[t]{0.25\textwidth}
    \centering
    \includegraphics[width=\textwidth]{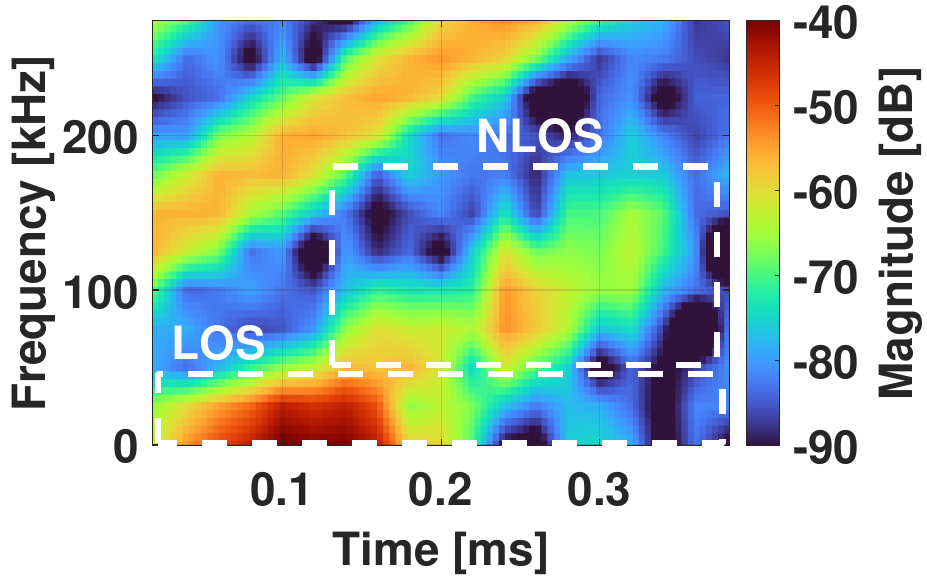}
    \caption{Multiplied signal}
    \label{fig:real_world_multiplied}
\end{subfigure}%
\begin{subfigure}[t]{0.25\textwidth}
    \centering
    \includegraphics[width=\textwidth]{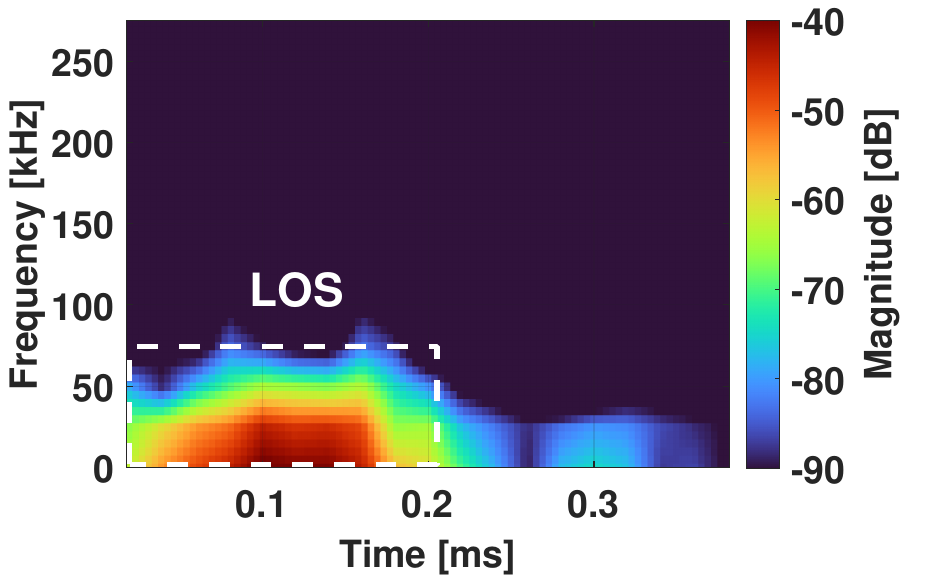}
    \caption{Filtered signal}
    \label{fig:real_world_filtered}
\end{subfigure}
\begin{subfigure}[t]{0.23\textwidth}
    \centering
    \includegraphics[width=\textwidth]{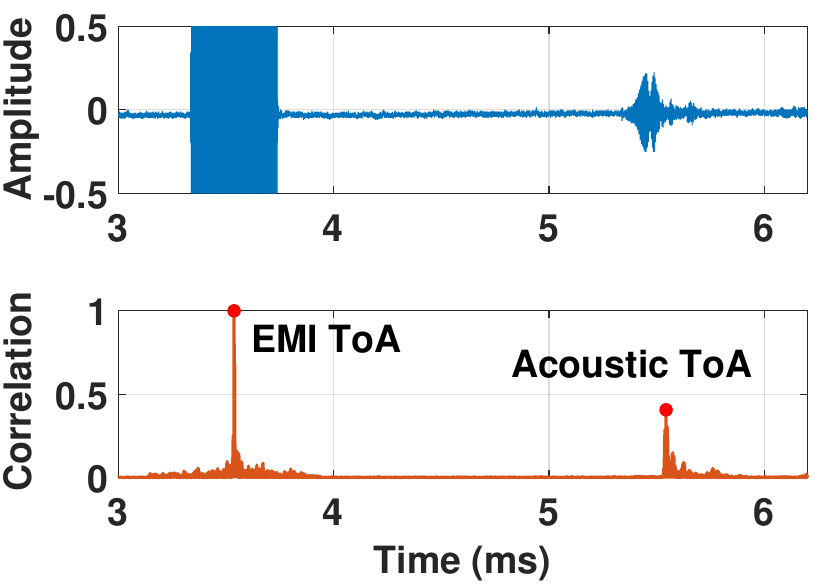}
    \caption{TDoA of EMI and Sound}
    \label{fig:EMI_toa}
\end{subfigure}\vspace{-5mm}
\caption{Received signal in the Real-World Environment. \textnormal{(a) Received chirp signal with a duration of 0.2 ms and a frequency of 125–250 kHz. (b) Signal after multiplication with the template chirp. (c) Filtered signal. (d) Estimated TDoA between the EM and acoustic signals.}}\vspace{-5mm}
\label{fig:real_world_siganl}
\end{figure*}

\noindent $\blacksquare$ \textbf{Multipath Challenges}.
The above AoA estimation method works well in environments without multipath. However, underwater environments exhibit a strong multipath effect, which severely distorts the spectrum. As shown in Fig.~\ref{fig:spectra_simulation}, the 0$^\circ$ spectrum in the simulated multipath channel is heavily distorted compared to the ideal spectrum without multipath.
The spectral distortion arises from the temporal overlap of the LOS and NLOS components as shown in Fig.~\ref{fig:real_world_raw_signal}. Next, we analyze the conditions under which temporal overlap occurs.

After propagation through the multipath underwater channel, the received signal can be written as
\begin{equation}\small
r(t)
= \alpha_{0}(t)s_{\theta}(t)+\sum_{k=1}^{M} \alpha_{k}(t)s_{\theta}\bigl(t - t_{k}\bigr),
  \label{eq:time_multipath}
  \end{equation}
where $i = 0, 1, 2, \ldots, M$ indexes each propagation path, $\alpha_i(t)$ denotes the attenuation along path $i$, and $t_k$ is the relative delay of the $k$-th NLOS path with respect to the LOS path ($i=0$).
Temporal overlap between the LOS and NLOS components occurs when the delay of the earliest NLOS path is shorter than the signal duration $T$, i.e.,
\begin{equation}
\min_{1 \le k \le M} t_k < T.
\end{equation}
\label{subsec:multipath_analysis}
\begin{figure}[t]
\centering
\begin{subfigure}[t]{0.23\textwidth}
    \centering
    \includegraphics[width=\textwidth]{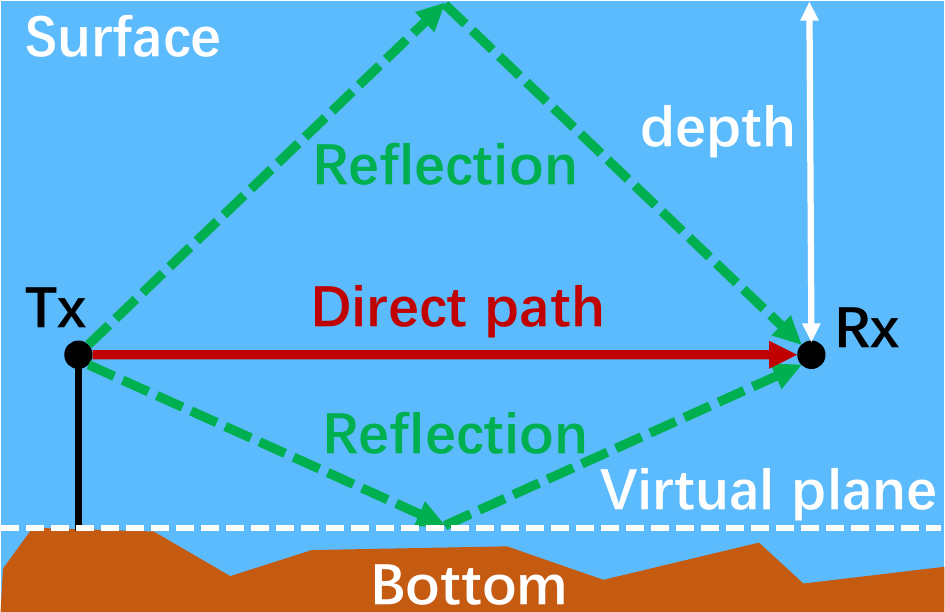}\vspace{-2mm}
    \caption{Multipath model}
    \label{fig:simulation_model}
\end{subfigure}
\begin{subfigure}[t]{0.24\textwidth}
    \centering
    \includegraphics[width=\textwidth]{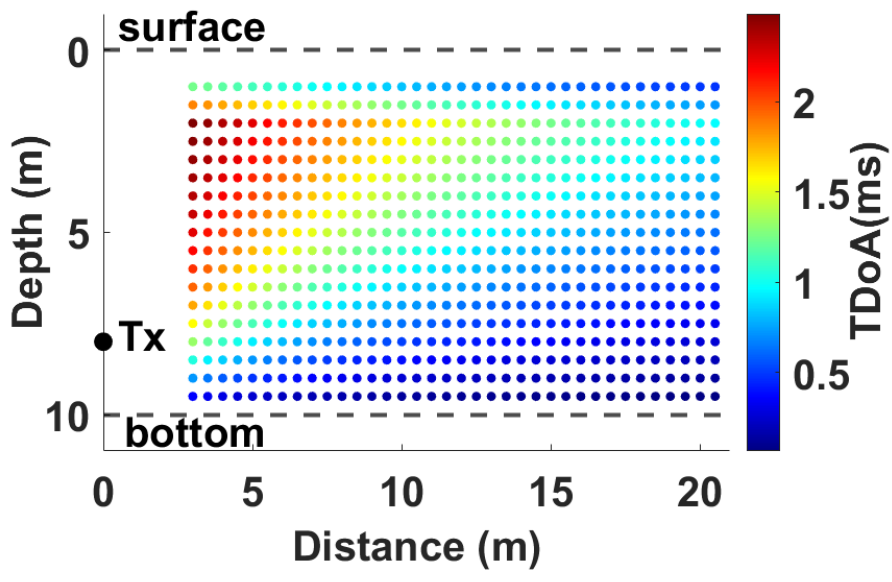}\vspace{-2mm}
    \caption{TDoA distribution}
    \label{fig:tdoa_distribution}
\end{subfigure}\vspace{-5mm}
\caption{Multipath Effect Simulation.\textnormal{(a) shows the theoretical propagation model in vertical plane. (b) shows the LOS-NLOS TDoA spatial distribution.}}\vspace{-5mm}
\label{fig:simulation}
\end{figure}
Therefore, knowing the minimal LOS-NLOS TDoA $\min t_k$ can help reduce multipath interference if we correctly choose the signal duration $T$ to be less than $\min t_k$. 

To acquire prior knowledge of the minimal LOS-NLOS TDoA. We employ the BELLHOP acoustic ray tracing tool~\cite{porter2011bellhop} to simulate the arrival time of every path's signal. 
In most practical underwater environments (e.g., oceans, lakes, and pools), the water depth is much smaller than the horizontal extent, so vertical reflections from the surface and bottom arrive much earlier than long-range horizontal reflections. We therefore model multipath propagation only in the 2D vertical plane, as illustrated in Fig.~\ref{fig:simulation_model}, because vertical paths yield a smaller LOS–NLOS TDoA than horizontal ones, providing a minimal estimate of the TDoA.
For irregular bottoms, accurately modeling bottom reflections is difficult, so we further approximate the environment by introducing a virtual horizontal reflector at the shallowest point of the bottom, which again gives the smallest possible LOS–NLOS TDoA.

For example, we simulate a 10m-deep underwater environment. The transmitter is fixed at a depth of 8 m, while the receiver is placed at depths ranging from 1 m to 9.5 m and at horizontal ranges from 3 m to 20 m relative to the transmitter. The TDoA between the LOS path and the earliest NLOS path is recorded.
The result is shown in Fig.~\ref{fig:tdoa_distribution}. We observe that as the receiver moves closer to the water surface or bottom, and as the horizontal distance from the transmitter increases, the TDoA decreases. Within the simulated region, the minimal TDoA is $0.065,\text{ms}$.
Through the simulation, we can estimate the minimal LOS-NLOS TDoA for a given environment. In principle, we can choose a chirp duration shorter than this minimal TDoA to avoid temporal overlap. However, the chirp duration $T$ cannot be arbitrarily reduced: making $T$ too short drastically reduces the total transmit energy, resulting in insufficient SNR for reliable long-range detection.
In our experiments, a chirp longer than $0.2\,\text{ms}$ is required for robust detection beyond $10\,\text{m}$. Consequently, in the scenario of Fig.~\ref{fig:tdoa_distribution}, a minimal TDoA of approximately $0.065\,\text{ms}$ makes LOS-NLOS overlap essentially unavoidable.
Once LOS and NLOS components overlap, separating them from the received waveform becomes a challenge. Next, we detail how to solve the overlap.

\noindent $\blacksquare$ \textbf{Multipath Suppression Algorithm.}
To address LOS-NLOS overlap, we propose a multipath suppression algorithm that attenuates the NLOS components in the received signal, thereby recovering a reliable directional spectrum from the LOS-only signal. The processed result is shown in Fig.~\ref{fig:lowpass_filterd_time_multipath}, We can show that the waveform with multipath effect is very similar to its ideal waveform. Next, we will detail the process of the algorithm.

First, different from analyzing the signal in the frequency domain as Eq.~\ref{eq:directional_signal_frequency}, we analyze the emitted signals' directional features in the time domain. 
The emitted signal $s_\theta(t)$ in $\theta$ can be expressed as 
\begin{equation}\small
  s_\theta(t)
  = \mathcal{F}^{-1}\!\bigl\{G_\theta(2\pi f)\,S(2\pi f)\bigr\}
  =
  g_\theta(t)\,
  \cos\!\Bigl(2\pi\bigl(f_0+\tfrac{k}{2}t\bigr)t\Bigr)
  \label{eq:chirp_aoa_response}
\end{equation}
where $g_\theta(t)$ denotes the direction-dependent envelope. By the stationary phase approximation~\cite{721375}, we obtain
\begin{equation}
    g_\theta(t)\approx G_\theta(kt+f_0)
    \label{eq:time_domain_features}
\end{equation}
which shows that $g_\theta(t)$ have the same waveform as the frequency response $G_\theta(f)$. Consequently, if we can recover $g_\theta(t)$ from the received signal with multipath interference, we can accurately estimate the AoA.

Next, we describe how to recover $g_\theta(t)$. 
When probing the chirp $r(t)$ depicted as Eq.~\ref{eq:time_multipath}
We multiply $r(t)$ by the original chirp in Eq.~\ref{eq:chirp} to obtain an product $m(t)$.
This product can be decomposed into three components: (i) a baseband term concentrated around zero frequency, corresponding to the LOS envelope; (ii) a term centered around the higher frequency $k t_k$, corresponding to NLOS interference; and (iii) a residual term with time-varying frequency components above $2f_0$. These three components are explicitly expressed as follow.
\begin{equation}\footnotesize
\begin{aligned}
m(t)
&= r(t)\cos\!\Bigl(2\pi(f_0+\tfrac{k}{2}t)t\Bigr)\\[2pt]
&= \alpha_0 s_\theta(t)\cos\!\Bigl(2\pi(f_0+\tfrac{k}{2}t)t\Bigr)\\
&\quad + \sum_{k=1}^{n} \alpha_k s_\theta(t-t_k)
      \cos\!\Bigl(2\pi(f_0+\tfrac{k}{2}t)t\Bigr)\\[4pt]
&=\underbrace{\frac{\alpha_0 g_\theta(t)}{2}}_{\text{LOS envelope}}+\underbrace{
    \sum_{k=1}^n 
    \frac{\alpha_k g_\theta(t-t_k)}{2}\,
    \cos(2\pi k t_k t + \beta_k)
   }_{\text{NLOS interference}} + \underbrace{h(t)}_{\text{residual term}}
\end{aligned}
\label{eq:signal_processing}
\end{equation}

where $\beta_{k}
= 2\pi\Bigl(\tfrac{k\,t_{k}^{2}}{2} - f_{0}\,t_{k}\Bigr)$ is the phase delay and
\begin{equation}\footnotesize
\begin{split}
h(t)=&\frac{\alpha_0\,g_\theta(t)}{2}\cos\bigl(2\pi(2f_0 + k\,t)\,t\bigr)\\
&+\sum_{k=1}^n\frac{\alpha_k\,g_\theta(t)}{2}
  \cos\!\Bigl(2\pi\bigl(2f_0 + k(t - t_k)\bigr)\,t
             +\frac{k\,t_k^2}{2}-f_0t_k\Bigr)
\end{split}
\label{eq:third_term}
\end{equation}
Fig.~\ref{fig:real_world_multiplied} shows the spectrogram of $m(t)$, where the LOS and NLOS components are clearly separated in frequency, with a spacing of approximately $k t_k$.

Then, to extract the LOS envelope near zero frequency while suppressing the NLOS component around $k t_k$ and the high-frequency term $h(t)$, we apply a low-pass filter with cut frequency 
\begin{equation}
    f_{cut}<kt_k
\end{equation}
to $m(t)$ and obtain the LOS component $n(t)$, whose spectrogram is shown in Fig.~\ref{fig:real_world_filtered}. and $n(t)$ can be approximated as
\begin{equation}
n(t)\approx \frac{\alpha_{0}\,g_{\theta}(t)}{2}
\end{equation}
Therefore, we extract the $g_\theta(t)$ from the received $r(t)$.

\textbf{Parameter Setting:} A discussion about how to set the parameters of signal and algorithm is detailed in Appendix~\ref{app:multipath}

\textbf{Vertical Direction Finding}: Although the AMS is not explicitly designed for elevation directivity, it naturally exhibits useful vertical directional patterns, as detailed in Appendix~\ref{app:depth_estimation}.

\subsection{Underwater Range Measurement}
\label{sec:ranging}
To recover the 3D position from a single anchor, we need to estimate the distance between the anchor and the receiver.
In this section, we propose an EM-acoustic TDoA ranging method, without requiring precise clock synchronization as classic schemes. 

In our scheme, we exploit the inherent electromagnetic leakage generated by the PZT transducer~\cite{gong2024enabling} as a timing reference. When the PZT is driven, it simultaneously emits an EM signal and an acoustic signal. The EM leakage propagates at approximately the speed of light, $c \approx 3\times 10^{8}\,\text{m/s}$, yielding an arrival time of
\begin{equation}
t_\text{EM} = \frac{20}{c} \approx 6.7\times 10^{-5}\,\text{ms}
\end{equation}
for a 20\,m path. In contrast, the acoustic signal propagates at the speed of sound in water, $v_\text{acoustic} \approx 1500\,\text{m/s}$, giving
\begin{equation}
t_\text{acoustic} = \frac{20}{v_\text{acoustic}} \approx 13.3\,\text{ms}.
\end{equation}
Since $t_\text{EM} \ll t_\text{acoustic}$, we can safely approximate $t_\text{EM} \approx 0$.
At the receiver, the ToA of the EM and acoustic signals are estimated via correlation-based detection, as illustrated in Fig.~\ref{fig:EMI_toa}. The propagation distance is then computed from the EM-acoustic time difference as
\begin{equation}
d = v_\text{acoustic} \bigl(t_\text{acoustic} - t_\text{EM}\bigr) \approx v_\text{acoustic}\, t_\text{acoustic},
\end{equation}
where $v_s$ is the speed of sound in water.

Furthermore, we analyze the attenuation of the EM signal. In our localization scheme, we use a chirp signal with a center frequency of $\sim$100~kHz, which lies in the low-frequency RF band. In fresh water, such low-frequency RF signals experience relatively weak attenuation~\cite{jiang2011electromagnetic}. For example, the attenuation of a 200~kHz RF signal is reported to be less than 8~dB over a 20~m path, indicating that these signals can reliably propagate over distances of several tens of meters.

\subsection{Scalable Underwater 3D Localization}
The previous sections described how to estimate the direction and range relative to an anchor. Building on these single-dimensional measurements, this section presents how the receiver estimates its 3D position in both single-anchor and multi-anchor systems. 

\noindent $\blacksquare$ \textbf{Single-Anchor Localization}.
When only a single anchor is available, we estimate the 3D position as follows.
For anchor $i$, we first compute the bearing angle $\theta_i$ as described in Sec.~\ref{sec:multipath}, and the range $r_i$ as described in Sec.~\ref{sec:ranging}. To recover the full 3D position, we further require the depth. We obtain the depth $z_i$ by running AoA estimation in the vertical plane as desctribed in Appendix ~\ref{app:depth_estimation}.

When the bearing angle $\theta_i$, a slant 3D range $d_i$, and a depth measurement $z_i$ is acquired. The 3D position $\hat{\mathbf{p}}^{(i)}$ is given as
\begin{equation}\small
  \hat{\mathbf{p}}^{(i)}
  = \mathbf{a}_i +
  \begin{bmatrix}
    h_i \cos\theta_i \\
    h_i \sin\theta_i \\
    z_i - a_{iz}
  \end{bmatrix},
\end{equation}
where $\mathbf{a}_i = (a_{ix}, a_{iy}, a_{iz})^\top$ is the position of anchor $i$ and the $h_i$ is the corresponding horizontal range that is calculate by $h_i = \sqrt{r_i^2 - (z_i - a_{iz})^2}$.

\noindent $\blacksquare$ \textbf{Multi-Anchor Localization}.
When measurements from multiple anchors are available, we estimate the 3D position by solving a weighted nonlinear least-squares problem.

We first define three per-anchor residuals.  
The bearing-angle residual for anchor $i$ is
\begin{equation}
\mathcal{L}_{\text{ang}}^{(i)} 
= \mathbf{n}_i^{\top}(\mathbf{p}-\mathbf{a}_i),
\label{eq:loss_bearing}
\end{equation}
where $\mathbf{p} = [x,y,z]^\top$ is the receiver position, $\mathbf{a}_i$ is the position of anchor $i$, and $\mathbf{n}_i = [-\sin\theta_i,\ \cos\theta_i,\ 0]^\top$
denotes the unit normal of the horizontal bearing line from anchor $i$.

The horizontal range residual is defined as
\begin{equation}
\mathcal{L}_{\text{rng}}^{(i)} 
= \rho_i(\mathbf{p}) - h_i,
\end{equation}
where$\rho_i(\mathbf{p})
= \bigl\|\,(x,y)^\top - (a_{ix}, a_{iy})^\top\bigr\|_2$
is the horizontal distance from anchor $i$ to $\mathbf{p}$, and $h_i$ is the measured horizontal range.

The depth residual is
\begin{equation}
\mathcal{L}_{\text{dep}}^{(i)} 
= z - z_i,
\end{equation}
where $z_i$ is the estimated depth relative to anchor $i$.

We then fuse all measurements by minimizing the following weighted sum of squared residuals:
\begin{equation}\small
\begin{aligned}
  \hat{\mathbf{p}}
  =&\arg\min_{\mathbf{p} \in \mathbb{R}^3}
  \sum_{i} w_i \Big(
    W_{\mathrm{ang}}\bigl(\mathcal{L}_{\text{ang}}^{(i)}\bigr)^2
  + W_{\mathrm{rng}}\bigl(\mathcal{L}_{\text{rng}}^{(i)}\bigr)^2
  + W_{\mathrm{dep}}\bigl(\mathcal{L}_{\text{dep}}^{(i)}\bigr)^2
  \Big),
\end{aligned}
\label{eq:localization}
\end{equation}
where $w_i$ is the per-anchor weight, and
$W_{\mathrm{ang}}, W_{\mathrm{rng}}, W_{\mathrm{dep}}$ are global weights
balancing the bearing, range, and depth residuals, respectively.

To avoid collisions among anchors during transmission, we coordinate the emitters using a classic TDMA scheme. Details are provided in Appendix~\ref{sec:TDMA}.

\begin{figure*}
    \centering
    \includegraphics[width=0.9\linewidth]{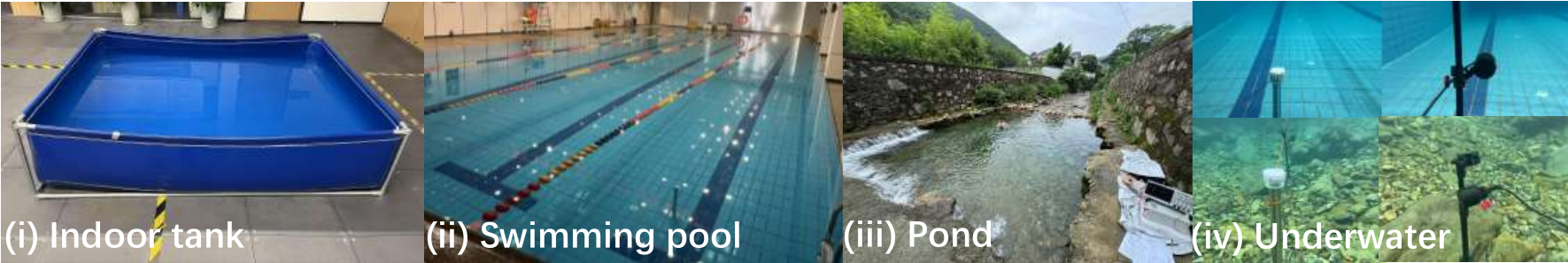}\vspace{-3mm}
\caption{Experimental Environments.
\textnormal{(i) Indoor water tank of $2\,\mathrm{m} \times 1.5\,\mathrm{m} \times 0.5\,\mathrm{m}$;
(ii) swimming pool of $25\,\mathrm{m} \times 12.5\,\mathrm{m} \times 1.5\,\mathrm{m}$;
(iii) outdoor pond of $10\,\mathrm{m} \times 10\,\mathrm{m}$ and a vary depth of $0.8\,\mathrm{m}$ to $1.8\,\mathrm{m}$;
(iv) underwater views of the anchors and receivers deployed in the swimming pool and the outdoor pond.}}\vspace{-3mm}
    \label{fig:scene}
\end{figure*}

\section{Implementation}
\label{sec:implementation}
\begin{figure}[t!]
    \centering
    \includegraphics[width=0.8\linewidth]{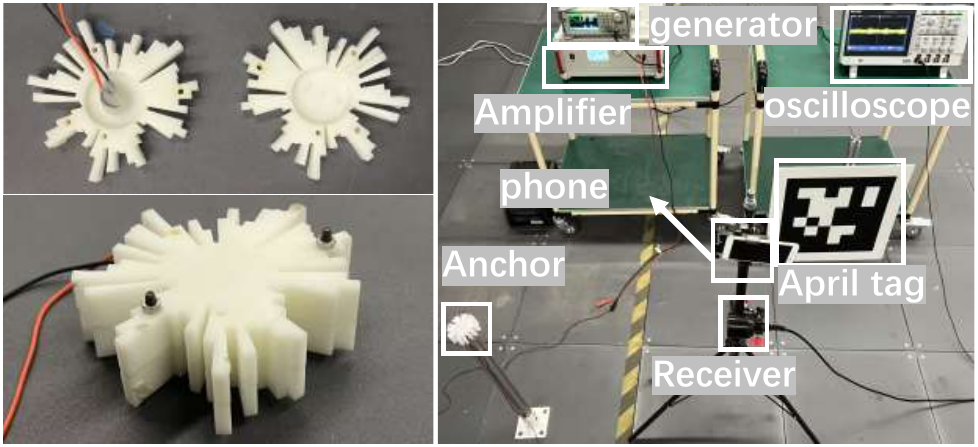}\vspace{-4mm}
    \caption{Implementation.
    }
    \vspace{-7mm}
    \label{fig:devices}
\end{figure}

This section details the implementation of \oursystem which consist of the anchor side and the receiver side.
The implementation is as Fig.~\ref{fig:devices}.

\textbf{Anchor}: 
Each anchor integrates a 3D-printed PLA AMS and a PZT transducer. The AMS is $24~\text{mm}$ thick and consists of annular sectors whose outer radii span $15$–$48\,\text{mm}$ according to the optimized design. A central cylindrical cavity (radius $15~\text{mm}$, height $10~\text{mm}$) houses the PZT, whose resonant frequency is $183~\text{kHz}$. The chirp waveform is generated by an SDG1022X signal generator~\cite{SiglentSDG1022XPlus} and amplified by an ATA-2088 high-voltage amplifier~\cite{AigtekATA2088} to a drive level of $400~\text{V}$. The chirp signals are tested in two bands, $125$–$375,\text{kHz}$ and $125$–$250,\text{kHz}$. \textit{Cost Analysis:} For each anchor, the 3D-printed AMS costs about \$6 and the PZT about \$5, for a total of roughly \$11, and the unit cost can be further reduced in large-scale production.

\textbf{Receiver}: A waterproof PZT transducer~\cite{DYW50200F} of resonant frequency of 200K is connected to an oscilloscope TektronixTBS2000B~\cite{TektronixTBS2000} is used as the receiver. 

\textbf{Dataset collection}:
\textit{For the spectral template in the calibration set:} we mount the anchor on a turntable, rotate it to predefined angles, and record the corresponding acoustic signals; the turntable angles serve as ground truth.
\textit{For localization test data:} we directly measure the $(x,y,z)$ coordinates by distance measurements. indoor distances are measured with a laser rangefinder relative to the walls, while outdoor distances we are measured by a marked rope.
\textit{For the moving-receiver demo} in Appendix~\ref{app:demo}, we attach an AprilTag to the receiver and use a smartphone camera to estimate its trajectory.

\section{Evaluation}
\begin{figure*}[t]
\centering
\begin{subfigure}[t]{0.235\textwidth}
    \centering    \includegraphics[width=\textwidth]{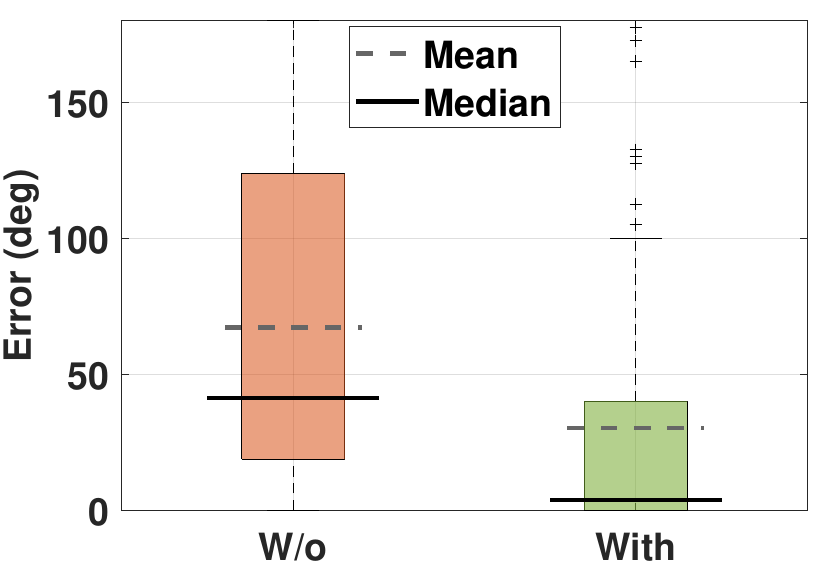}\vspace{-3mm}
    \caption{AMS}
\label{fig:with_wo_AMS}
\end{subfigure}
\begin{subfigure}[t]{0.235\textwidth}
    \centering    \includegraphics[width=\textwidth]{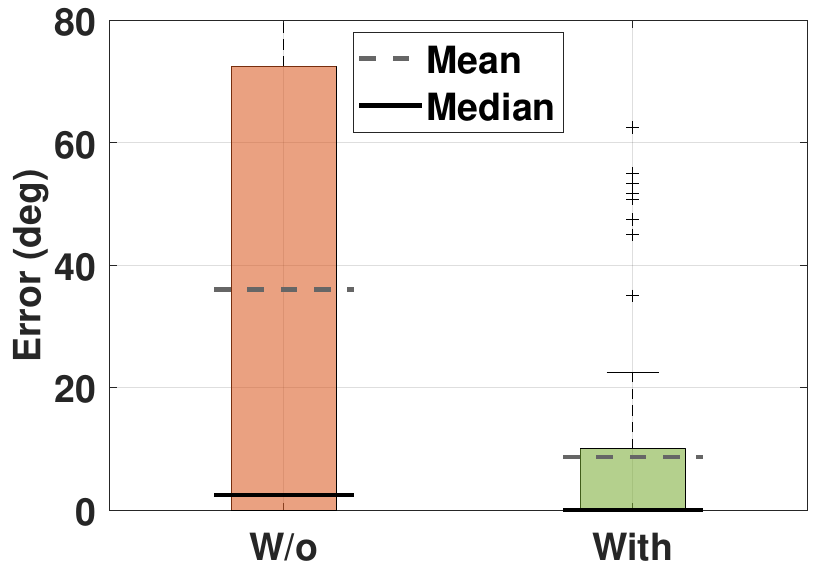}\vspace{-3mm}
    \caption{Multipath suppression}
\label{fig:error_filtered_ablation}
\end{subfigure}
\label{fig:error_aoa}
\begin{subfigure}[t]{0.25\textwidth}
    \centering    \includegraphics[width=\textwidth]{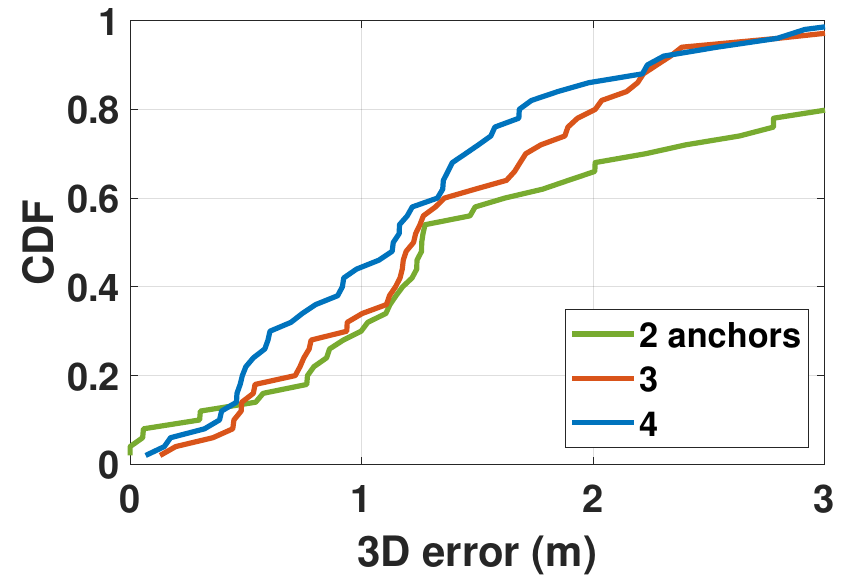}\vspace{-3mm}
    \caption{AoA-only}
\label{fig:3D_error_only_aoa}
\end{subfigure}%
\begin{subfigure}[t]{0.25\textwidth}
    \centering    \includegraphics[width=\textwidth]{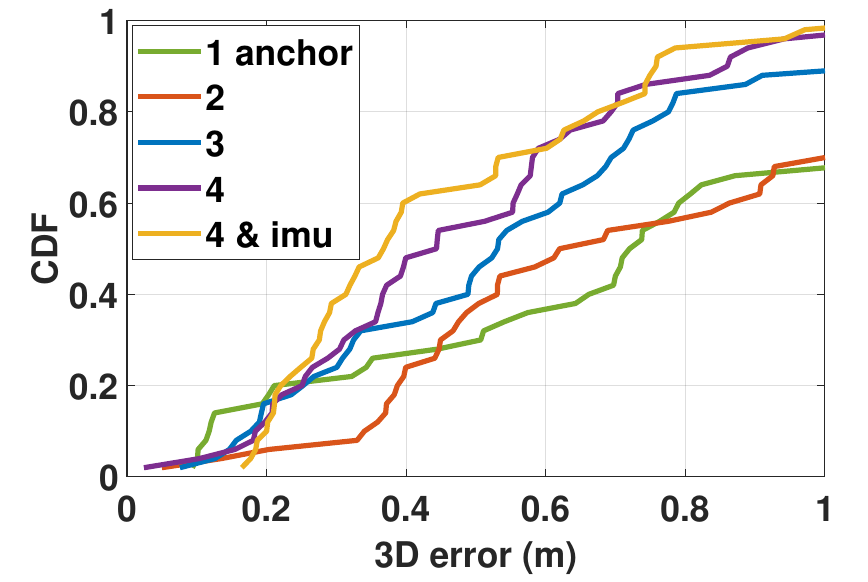}\vspace{-3mm}
    \caption{AoA-Ranging joint}
\label{fig:3D_error_joint_localization}
\end{subfigure}
\vspace{-5mm}
\caption{Evaluation Results. \textnormal{(a) and (b) illustrate the AoA estimation errors in the ablation studies of the AMS and the multipath suppression algorithm while (c) and (d) illustrate the localization errors in AoA-only localization and joint AoA–ranging localization.}}\vspace{-5mm}
\label{fig:error_aoa}
\end{figure*}

\subsection{Microbenchmark}
\noindent $\blacksquare$ \textbf{Experiment Setup}. To evaluate the effectiveness of the proposed AMS in Sec~\ref{sec:metasurface} and multipath suppression algorithm in Sec.~\ref{sec:multipath}, we conduct experiments to evaluate the AoA estimation in an indoor water tank as depicted in Fig.~\ref{fig:scene} (i).
The anchor and the receiver are at a depth of $15\, \text{cm}$ and have a varying anchor-receiver range.
The emitted chirps have a duration of $0.2\,\text{ms}$ and a bandwidth of 250\,kHz. The receiver's sample frequency is 31.25 MHz.
The dataset is collected over a full $360^\circ$ in azimuth at different anchor-to-receiver distances.
Simulation results indicated that the LOS-NLOS TDoA is from $0.04\,\text{ms}$ to $0.09\,\text{ms}$, which is significantly shorter than the $0.2\,\text{ms}$ duration, resulting in $55\%$--$80\%$ temporal overlap.

\noindent $\blacksquare$ \textbf{Validation of the AMS Effectiveness.}
To verify the AMS's effectiveness in generating the directional feature.
Two AoA estimation experiments are conducted under settings with and without AMS. 
We use the data collected in the anchor-receiver range of $0.4\,\text{m}$ as the spectral template. The test set is in the anchor-receiver range of $0.3\,\text{m}$. We use the raw spectral data for AoA estimation.  
Figure~\ref{fig:with_wo_AMS} shows the result. Without AMS, the AoA estimation yields a mean angular error of 67.1$^{\circ}$. This substantial error indicates that the estimation can resolve only very coarse directions, because the PZT itself provides insufficient directional features. However, integrating the AMS reduces the mean error to 30.2$^\circ$, corresponding to a 55.0\% improvement under identical settings. This improvement shows that more features are embedded with AMS.

\noindent $\blacksquare$ \textbf{Multipath Suppression Algorithm.}
We then test the effectiveness of our proposed multipath suppression algorithm.
We use the datasets collected at anchor-receiver distances of 0.4\,m, 0.5\,m, 0.6\,m, and 0.7\,m. 
we use the data of 0.4\,m and 0.7\,m with a 2.5$^\circ$ spacing in 0-360 $^\circ$ as the template. and use the data of 0.5\,m and 0.6\,m with a 5$^\circ$ spacing in 0-360 $^\circ$ as the test data. We first use the raw spectrum in the 125-375\,kHz band for AoA estimation as the baseline. 
And then use the algorithm-processed signal for AoA estimation. As shown in Fig.~\ref{fig:error_filtered_ablation}, the raw spectrum predictor yields a mean AoA estimation error of 36.1$^{\circ}$.
The significant error here comes from spectral distortion caused by multipath effects as we describe in~\ref{subsec:multipath_analysis}. 
After the algorithmic process, the error reduces to 8.7$^{\circ}$, yielding a 77.6\% improvement. The improvement shows the validity of the proposed algorithm. It is worth noting that the template and the test data are collected in different ranges and remain accurate for AoA estimation. This means we only need to collect data in a few selected calibration ranges, and the remaining ranges of the tested data can also be accurately estimated.
\subsection{Localization Accuracy}
\noindent $\blacksquare$ \textbf{Experiment Setup}
The localization experiments are conducted in the swimming pool as Fig~\ref{fig:scene}(ii). 4 anchors are deployed at the four corners of an 8 m $\times$  8 m square at a depth of 0.8m.
A 0.4 ms chirp spanning 125-375 kHz is used as the probing signal. Simulation shows that the time TDoA between the direct path and the first indirect path is $0.05\,\mathrm{ms}$ to $0.38\,\mathrm{ms}$ (anchor-receiver distance 1 to 10 m).
 Accordingly, a low-pass filter with a cutoff frequency of 70 kHz is applied to suppress the multipath effect and improve the robustness of TDoA estimation. 
For each anchor, training data are collected at ranges of $1$, $3$, $5$, and $7\,\mathrm{m}$ with an angular spacing of $5^{\circ}$. Data in depths of $0.5\,\mathrm{m}$ and $0.8\,\mathrm{m}$ are collected. For evaluation, we select 50 test locations approximately uniformly distributed within the square region, and at each location we record signals from all four anchors.

\noindent $\blacksquare$ \textbf{AoA-Only Localization}
First, we use naive multiple anchors' AoA information for localization. Each anchor first estimates the azimuth $\theta$ and depth $z$. After acquiring the estimates from multiple anchors, the $(x,y,z)$ coordinates are obtained via optimization. As shown in Fig.~\ref{fig:3D_error_only_aoa}, the median 3D errors with 2, 3, and 4 anchors are $1.26,\mathrm{m}$, $1.22,\mathrm{m}$, and $1.15,\mathrm{m}$, respectively.
This result shows that our naive AoA-based localization achieves moderate accuracy, comparable to recent acoustic AoA and ToA methods, as shown in Table~\ref{tab:comparison}.
We then analyze the error as a function of the number of anchors. Relative to the 2-anchor baseline, the median error of 3- and 4-anchor improves by $3.17\%$ and $8.73\%$. The improvements in median error as a function of the number of anchors are moderate.
Further, the corresponding $75^{\text{th}}$-percentile errors are $2.74,\mathrm{m}$, $1.88,\mathrm{m}$, and $1.57,\mathrm{m}$, representing improvements of $31.39\%$ and $42.70\%$ for 3- and 4-anchor. The improvements are obvious.
This indicates that, for samples with minor errors, two anchors already provide accurate localization for many samples, and additional anchors offer only limited average gains. However, for samples with significant errors, adding more anchors rapidly shrinks these outliers, leading to pronounced improvements in upper-tail error metrics.

\noindent $\blacksquare$ \textbf{AoA-Ranging Joint Localization} We further evaluate 3D localization performance using the AoA-ranging joint method, where the distances between the receiver and anchors are obtained from ToA measurements of EM and acoustic signals.
As shown in Fig.~\ref{fig:3D_error_joint_localization}, the median 3D errors with 1, 2, 3, and 4 anchors are $0.73~\mathrm{m}$, $0.65~\mathrm{m}$, $0.53~\mathrm{m}$, and $0.44~\mathrm{m}$, respectively.
Compared to the single-anchor baseline, the improvements are $16.4\%$, $27.4\%$, and $39.7\%$ for 2, 3, and 4 anchors. The corresponding  $75^{\text{th}}$-percentile errors are $1.25~\mathrm{m}$, $1.05~\mathrm{m}$, $0.72~\mathrm{m}$, and $0.63~\mathrm{m}$, yielding improvements of $16.0\%$, $42.4\%$, and $49.6\%$. 
These results suggest that multiple anchors substantially increase the localization accuracy. 
The first reason is that more anchors introduce additional independent constraints and redundant observations. All anchors jointly determine the position, so random noise is averaged out, and a few outliers are diluted by the majority of reliable measurements.
The second reason is the improved spatial coverage. The localization accuracy of a single anchor is high at short range. Still, it degrades with distance (e.g., for single-anchor accuracy, the median error is 0.54 m within 0-6 m but increases to 1.22 m within 6-10 m). Therefore, deploying more anchors ensures that more points fall within the high-accuracy range of nearby anchors, substantially improving overall localization performance.

\noindent $\blacksquare$ \textbf{Fusing IMU data via Kalman Filter}
The IMU sensor is widely used in underwater robots. We evaluate localization accuracy when fusing IMU data. Using the MATLAB imuSensor tool~\cite{matlab-imusensor}, we synthesized IMU data with realistic noise and bias drift, closely resembling real sensor measurements. As shown in Fig.~\ref{fig:3D_error_joint_localization}, by integrating this data via a Kalman filter, we reduced the median localization error in the 4-anchor setup from 0.44 m to 0.37 m, achieving a 15.9\% reduction in overall positioning error. 
\begin{figure*}[t]
\centering
\begin{minipage}[t]{0.25\textwidth}
    \centering
    \includegraphics[width=\linewidth]{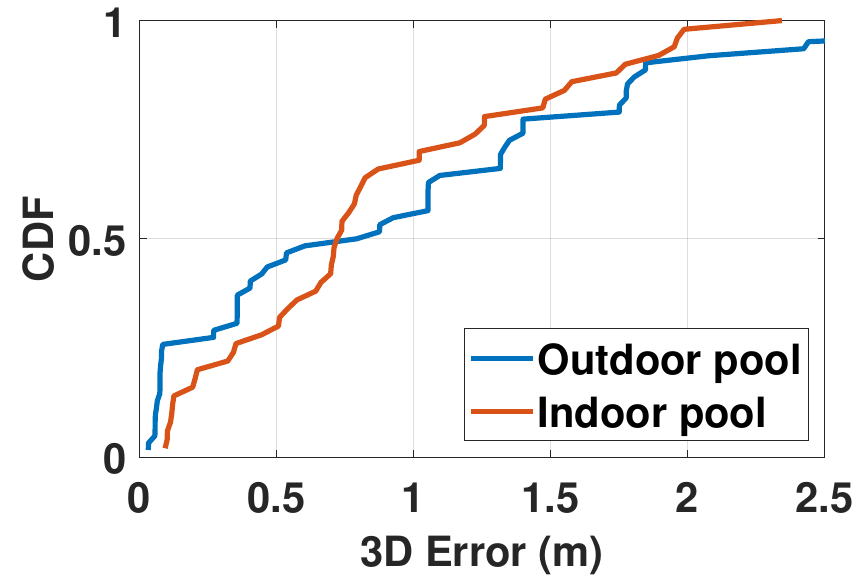}\vspace{-3mm}
    \caption{Different Scenes.}
    \label{fig:localization_scene}
\end{minipage}
\begin{minipage}[t]{0.24\textwidth}
    \centering    \includegraphics[width=\textwidth]{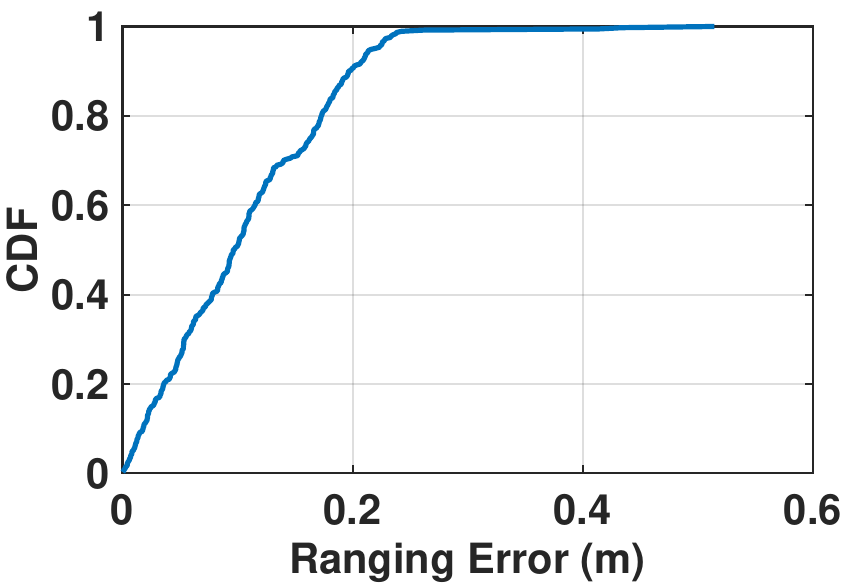}\vspace{-3mm}
    \caption{Ranging.}
\label{fig:ranging_error}
\end{minipage}
\begin{minipage}[t]{0.24\textwidth}
    \centering    \includegraphics[width=\textwidth]{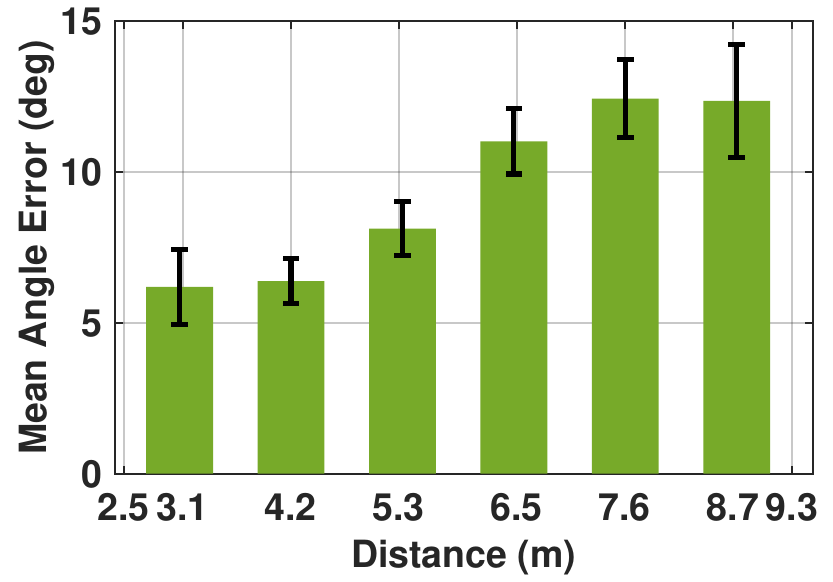}\vspace{-3mm}
    \caption{AoA Error}
\label{fig:aoa_error_vs_distance}
\end{minipage}
\begin{minipage}[t]{0.24\textwidth}
    \centering    \includegraphics[width=\textwidth]{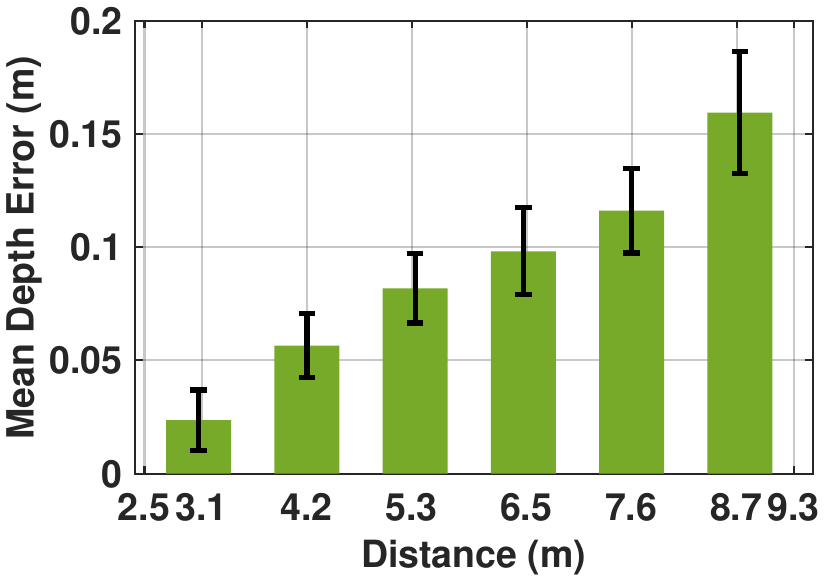}\vspace{-3mm}
    \caption{Depth Error.}
\label{fig:depth_error_vs_distance}
\end{minipage}
\vspace{-5mm}
\label{fig:setting_influence}
\end{figure*}
\subsection{Impact Analysis}
\noindent $\blacksquare$ \textbf{Different Scenes}
To evaluate the robustness of our method under different environmental conditions. We conduct an extra experiment in an outdoor pond as Fig~\ref{fig:scene}(iii). A total of 62 test samples are collected at distances of 2-6 m, angular ranges of 0-90 $^\circ$, and depths of 0.5 m and 0.8 m to evaluate localization accuracy.
As shown in Fig.~\ref{fig:localization_scene}, the 3D localization accuracy achieves a median error of 0.83 m, a $75^{\text{th}}$-percentile error of 1.40 m, and a $90^{\text{th}}$-percentile error of 1.91 m. 
Compared with indoor experiments, the accuracy degrades slightly: the median error is 0.83 m within 0-6 m in the wild, versus 0.73 m within 0-9.3 m indoors. This degradation mainly arises from irregular, shallow bathymetry (local depths down to 0.8 m) and obstacles such as rocks, which create strong NLOS paths whose LOS-NLOS TDoAs are very short, reducing the accuracy. Nevertheless, the field results still demonstrate that the proposed method maintains sub-meter accuracy in wild environments.

\noindent $\blacksquare$ \textbf{Ranging Accuracy}
We evaluate ranging accuracy using a dataset collected from 0 to 9.3 m. As shown in Fig.~\ref{fig:ranging_error}, the median error is 0.097 m, and 90\% of the measurements have an error below 0.198 m. These results show that the proposed EM-acoustic TDoA method achieves reliable underwater ranging performance without requiring time synchronization.

\noindent $\blacksquare$ \textbf{AoA Accuracy vs Distance}
We evaluate AoA estimation accuracy as a function of distance using a dataset collected from 0 to 9.3 m.
As shown in Fig.~\ref{fig:aoa_error_vs_distance}, the mean AoA estimation error clearly increases with distance, rising from 6.18$^\circ$ at 2.5-3.6 m to 12.34$^\circ$ beyond 8 m. This degradation is primarily attributed to the attenuation of acoustic signal strength with distance, which reduces the effective SNR and weakens the direction-dependent spectral features. Consequently, the classifier becomes less discriminative, leading to larger AoA deviations at longer ranges.

\noindent $\blacksquare$ \textbf{Depth Estimation Error}
We evaluate depth estimation accuracy as a function of distance using a dataset collected from 0 to 9.3 m.
As shown in Fig.~\ref{fig:ranging_error}, the mean depth estimation error clearly increases with distance, which grows from 0.02 m in the nearest bin to 0.16 m as the distance increases. Two major factors contribute to this increase: First is the reduced signal SNR as distance increases. Second is the elevation angles $\arctan(\frac{d_1-d_2}{distance})$ relative to the anchors decrease as distance increases (5.71$^\circ$ in 3 m and 1.91 $^\circ$ in 9 m), resulting in less distinction in the spectrum feature. 

\noindent $\blacksquare$ \textbf{Accuracy in Decoding}
Subsequently, we evaluate the decoding performance of our system. We adopt a binary modulation scheme, encoding the presence of the chirp as ‘1’ and its absence as ‘0’. Fig.~\ref{fig:amplitude_vs_distance} reports the bit error rate (BER) as a function of distance. Our system exhibits strong decoding robustness: the BER remains below 0.8\% for distances between $1$ and $10,\mathrm{m}$, and stays within 6\% even in the challenging $10$--$12,\mathrm{m}$ range. 

\noindent $\blacksquare$ \textbf{Impact of Signal Bandwidth.}  
We evaluate the AoA estimation error as a function of signal bandwidth. We use the datasets collected at 0.4\,m (training) and 0.3\,m (testing). Two bandwidth settings are compared: a narrow band of 125\,kHz (125–250\,kHz) and a wide band of 250\,kHz (125–375\,kHz). A spectrum-based estimator is used for AoA estimation. 
The narrow‑band configuration yields a mean absolute AoA error of 43.4$^\circ$, whereas the wide‑band configuration reduces the error to 30.2$^\circ$. This corresponds to a 30.4\% relative improvement, demonstrating that broader bandwidth provides richer spectral features and thus enhances directional estimation.\vspace{-5mm}
\subsection{Transmission Efficiency}
\begin{figure*}[t]
\centering
\begin{minipage}[t]{0.26\textwidth}
    \centering    \includegraphics[width=\textwidth]{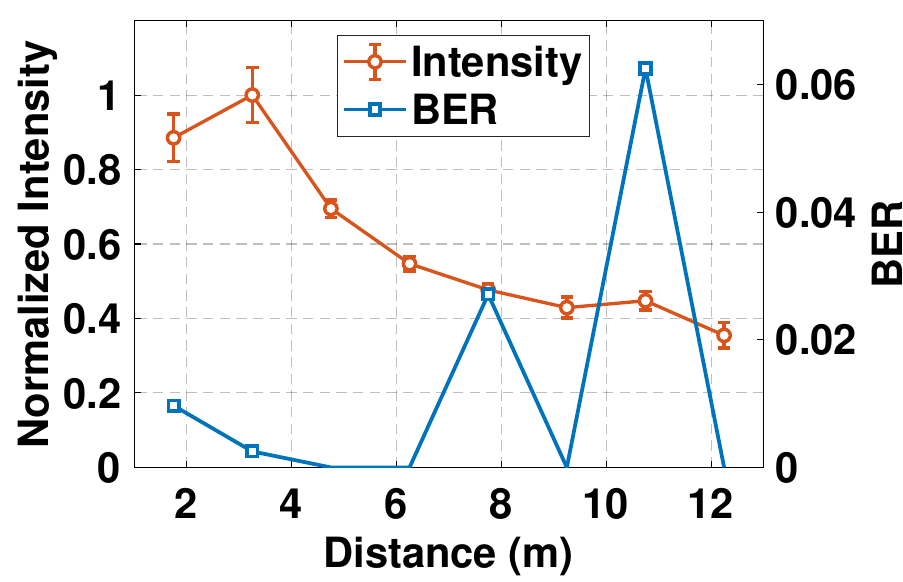}\vspace{-3mm}
    \caption{Intensity and BER.}
\label{fig:amplitude_vs_distance}
\end{minipage}
\hfill
\begin{minipage}[t]{0.24\textwidth}
    \centering 
    \includegraphics[width=\textwidth]{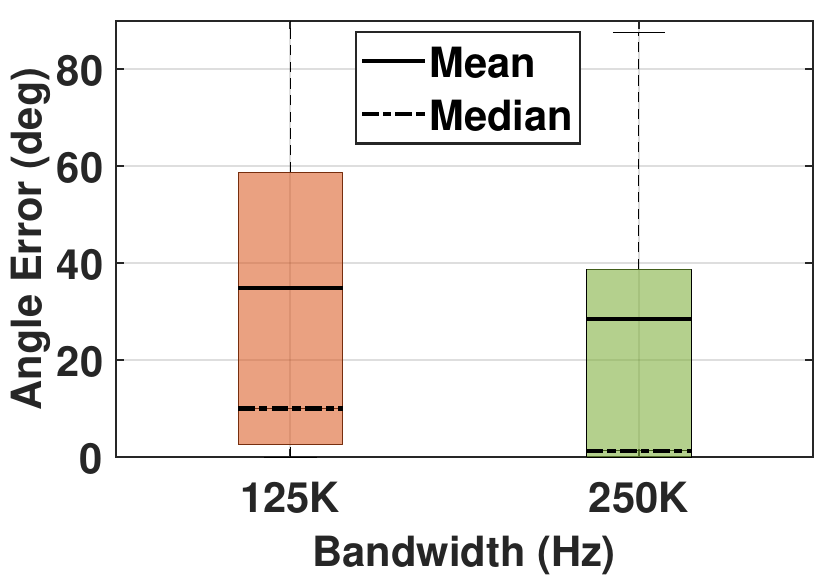}\vspace{-3mm}
    \caption{Bandwidths.}
\end{minipage}
\hfill
\begin{minipage}[t]{0.25\textwidth}
    \centering    \includegraphics[width=\textwidth]{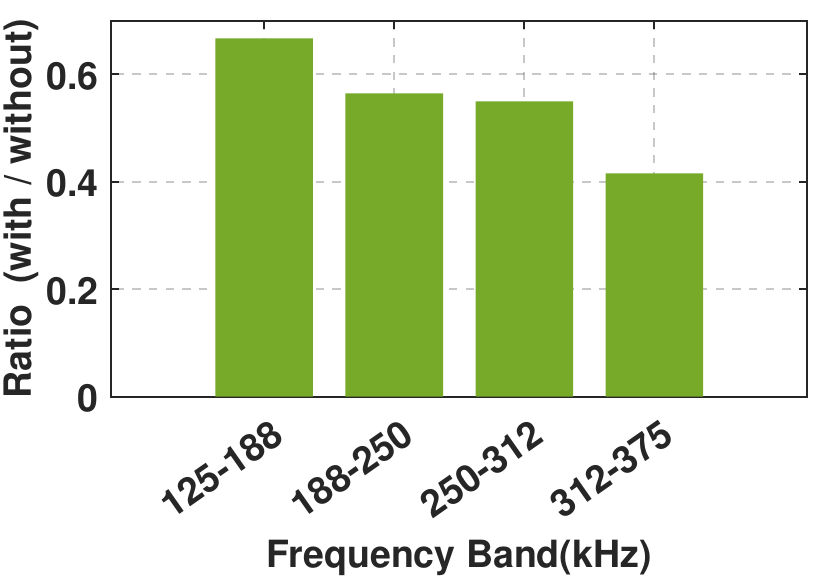}\vspace{-3mm}
    \caption{Transmission.}
\label{fig:amplitude_vs_frequency}
\end{minipage}
\hfill
\begin{minipage}[t]{0.2\textwidth}
    \centering    \includegraphics[width=\textwidth]{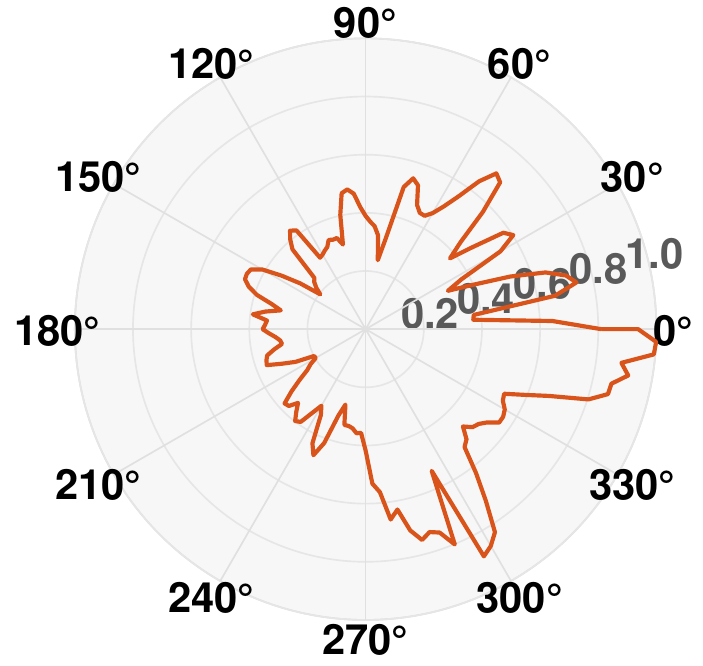}\vspace{-3mm}
    \caption{Amplitude.}
\label{fig:amplitude_vs_angle}
\end{minipage}\vspace{-5mm}
\end{figure*}

\noindent $\blacksquare$ \textbf{Transmission Efficiency vs Frequency}
We quantify the overall amplitude transmission coefficient as the ratio between the received signal amplitude with the AMS and that without the AMS, averaged over the full 0$^\circ$-360$^\circ$ angular range. The resulting mean amplitude transmission coefficient is 0.63.
Besides, we test the transmission as a function of the frequency. We divide the frequency band into four segment and calculate the ratio respectively.
Fig.~\ref{fig:amplitude_vs_angle} shows that the amplitude ratio between the AMS-assisted and non-AMS configurations decreases monotonically with increasing frequency, from 0.67 in the 125–188 kHz band to 0.42 in the 312–375 kHz band. The result demonstrate the attenuation fo unit 3D-printed AMS is higher as the frequency increase. 

\noindent $\blacksquare$ \textbf{Intensity in different directions}
We calculate the signal intensity at angles from $0^\circ$ to $360^\circ$ with a spacing of $2.5^\circ$.
As shown in Fig.~\ref{fig:amplitude_vs_angle}, the normalized intensity exhibits significant variation. The intensity diversity arises from the different thicknesses of the AMS, which lead to different attenuation. Second is that the far-field sound pressure has diversity because of the modulation of AMS.

\section{Related works}
We detail the prior works about underwater localization and acoustic metasurface. And some representive systems are summarized in the Table.~\ref{tab:comparison}.
\subsection{Underwater Localization}
Among underwater localization systems for autonomous underwater vehicles (AUVs) and divers, many approaches have been explored, including IMU-based~\cite{li2019multiple,salavasidis2019terrain}, visible-light-based~\cite{zhang2022u,lin2024uwbeacon}, RF-based~\cite{park20163d}, and acoustic methods~\cite{webster2009preliminary,kebkal2017underwater,liu2009linear,syed2006time,chirdchoo2008mu,lu2010d,chen2023underwater,lee2009novel,tan2010cooperative,huang2018node,afzal20243d}. Li et al.~\cite{li2019multiple} propose an IMU-based dead-reckoning framework for AUV localization, which periodically relies on acoustic updates from to correct IMU drift; in general, IMU-only localization suffers from rapid error accumulation and must be fused with other signals for acceptable accuracy. For visible-light methods, U-Star~\cite{zhang2022u} uses passive tags as anchors and UWBeacon~\cite{lin2024uwbeacon} uses LEDs as beacons, both achieving good performance in clear water but degrading under turbid or low-light conditions. Park et al.~\cite{park20163d} estimate distance from received RF signal strength, yet RF waves experience severe attenuation within several decimeters underwater, making them unsuitable for long-range localization.

Acoustic underwater localization methods mainly include time-of-arrival (ToA), angle-of-arrival (AoA), and fingerprinting. ToA-based approaches~\cite{chen2023underwater,lee2009novel,webster2009preliminary,kebkal2017underwater,liu2009linear,syed2006time,chirdchoo2008mu,lu2010d} require precise time synchronization, relying on either expensive high-precision clocks or complex message-exchange protocols. AoA-based systems~\cite{huang2018node,afzal20243d} depend on hydrophone arrays, increasing hardware size, cost, and power consumption. Fingerprint-based methods~\cite{lee2009novel,tan2010cooperative} demand large offline datasets, which are hard to collect underwater. As a result, existing methods suffer from hardware cost, synchronization overhead, or data scarcity, whereas \oursystem achieves underwater localization with a lightweight, scalable, synchronization-free design.

\subsection{Acoustic Metasurface}
Acoustic metasurfaces are powerful tools for manipulating sound waves, enabling customized acoustic field distributions through carefully designed subwavelength structures such as coiling-up spaces~\cite{memoli2017metamaterial} and Helmholtz resonators~\cite{fang2006ultrasonic}. They have been widely used to enhance the performance of acoustic communication~\cite{zhang2023acoustic}, imaging~\cite{bai2022spidr}, and localization systems~\cite{wang2025metasonic}. For localization, Owlet~\cite{garg2021owlet} designs a structure combining subwavelength apertures to enable AoA estimation, while MetaAng~\cite{fu2024pushing} and MetaSonic~\cite{wang2025metasonic} employ curled-pillar-based AMS designs for AoA estimation. However, all of these systems operate in airborne environments.

Underwater acoustic metamaterial have been extensively explored for applications such as invisibility cloaking~\cite{zhang2011broadband,chen2017broadband,zhou2022underwater}, beamforming~\cite{zhang2019subwavelength,jimenez2019generating}, and acoustic absorption~\cite{zhou2023ultrathin,qu2022underwater}.
However, only a few studies investigate AMS-based underwater localization. Among them, we focus on the works most closely related to directional sensing.
Jin et al.~\cite{jin2024bubble} introduced a bubble-based metamaterial that enhances acoustic amplitude in specific directions. Li et al.~\cite{li2024metastructure} proposed a hemispherical metastructure composed of serially connected pore-cavity units, mounted on the hydrophone side for directional detection. However, the theoretical analysis of AMS is lacking, so we cannot optimize its parameters. Moreover, its structure is more complex to produce than that of our AMS.
Chen et al.~\cite{chen2019broadband} further designed a metasurface to convert underwater cylindrical waves into plane waves, but it is designed for wavefront shaping rather than localization. However, the AMS is made of metal, which is costly.

\section{Conclusion}
We propose \oursystem, the first metasurface-assisted underwater robot localization system that embeds direction-dependent acoustic features directly into wideband signals. By jointly designing a waterborne acoustic metasurface, a spectral-feature-based AoA estimation algorithm, and a multipath-resilient signal processing pipeline, \oursystem \\overcomes long-standing challenges of impedance matching and severe underwater multipath. Our approach introduces a new paradigm of hardware–algorithm co-design for underwater sensing, enabling compact hardware, synchronization-free operation, and sub-meter 3D localization accuracy across diverse aquatic environments. More broadly, \oursystem points toward a future where intelligent acoustic metamaterials empower scalable, low-cost, and high-precision localization for underwater robots, environmental monitoring, and large-scale aquatic IoT systems.

\bibliographystyle{ACM-Reference-Format}
\bibliography{related_works}

\appendix
\section{Appendix}
\subsection{Example of the AMS optimization}
\label{app:optimization}
To solve the optimization problem as ~\ref{eq:optimization}, We first discretize the angular domain into $M$ directions $\{\beta_1,\beta_2,\ldots,\beta_M\}$ and the frequency domain into $L$ bins $\{f_1,f_2,\ldots,f_L\}$. For each direction $\beta_i$, we compute a spectral amplitude vector $v_i$ according to Eq.~\ref{eq:far_field}. The spectral similarity $G_{i,j}$ between directions $\beta_i$ and $\beta_j$ is then quantified using cosine similarity:
\begin{equation}\small
G_{i,j} = \frac{\|v_i \cdot v_j\|}{\|v_i\| \,\|v_j\|},
\label{eq:cosine_similarity}
\end{equation}
where a value close to $1$ indicates highly similar and to $0$ indicates low similarity.

Specifically, we set $N = 60$, a maximum thickness of $D = 3.3~\text{cm}$, The angular domain is discretized into 360 directions with a $1^{\circ}$ resolution, and the frequency domain into 101 bins from 100~kHz to 200~kHz. By solving the optimization problem in Eq.~\ref{eq:optimization}, the mean similarity between any two directions is reduced from 0.87 for a random configuration to 0.75 for the optimized configuration.

\subsection{Scaling up by Coarse TDMA}\label{sec:TDMA}
When deploying multiple anchors, distinguishing between them is crucial. We use the TDMA method, which divides time into multiple slots, with each anchor occupying a unique slot. In practice, all anchors will roughly synchronize themselves using the network synchronization service and then broadcast their signals in a predefined time order and slots to prevent interference. Each signal contains three fields: Preamble, ID, and Guard Interval. The preamble is a 0.4 ms chirp signal for localization and alignment with the ID field. The ID field contains 8 bits, with each bit lasting 0.2 ms. These bits are encoded using FM0, with the final bit reserved for parity checking. This encoding scheme supports up to 128 anchors. A 0.2 ms guard interval is included to prevent interference from echoes due to synchronization errors and sound delay. Overall, the whole signal is a 2.2~ms signal.
\subsection{Discussion: Parameters in Multipath}
\label{app:multipath}
\begin{figure}[h!]
\centering
\begin{subfigure}[t]{0.24\textwidth}
    \centering    \includegraphics[width=\textwidth]{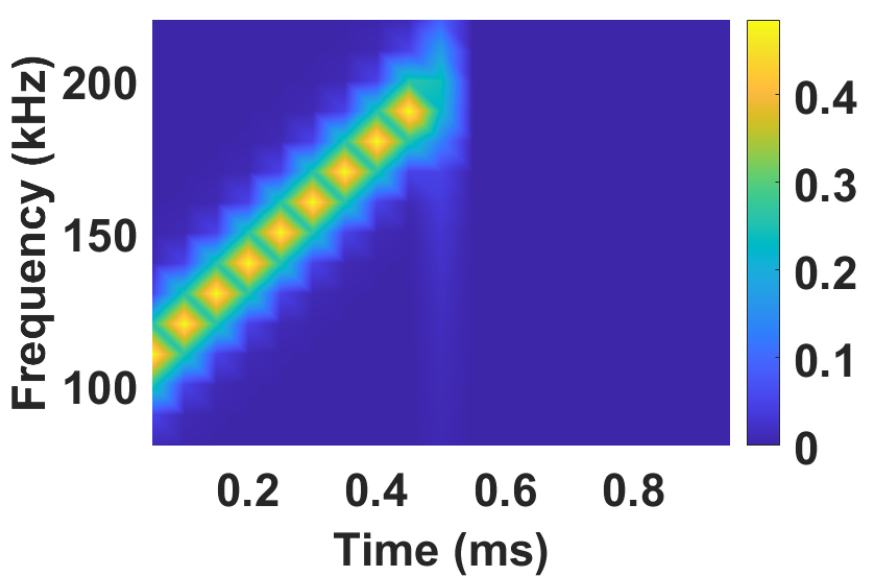}\vspace{-3mm}
    \caption{Transmitted chirp}
\label{fig:original_signal}
\end{subfigure}%
\begin{subfigure}[t]{0.24\textwidth}
    \centering
    \includegraphics[width=\textwidth]{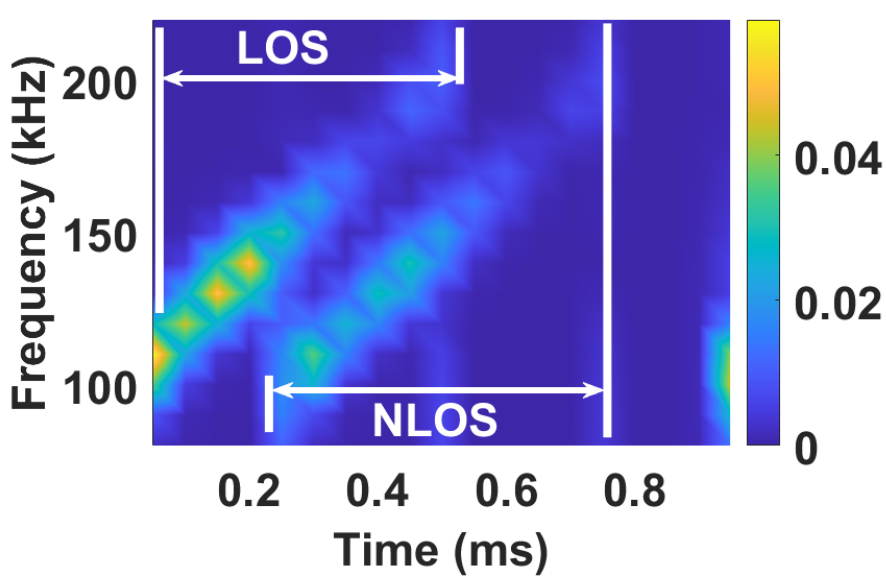}\vspace{-3mm}
    \caption{Received signal}
    \label{fig:original_multipath}
\end{subfigure}
\begin{subfigure}[t]{0.24\textwidth}
    \centering
    \includegraphics[width=\textwidth]{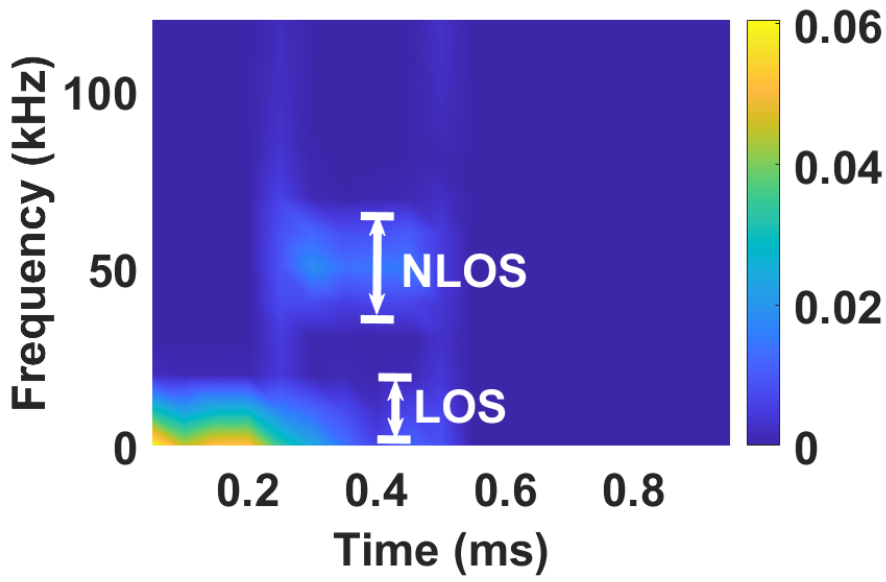}\vspace{-3mm}
    \caption{Multiplication with chirp}
    \label{fig:multipled_multipath}
\end{subfigure}%
\begin{subfigure}[t]{0.24\textwidth}
    \centering
\includegraphics[width=\textwidth]{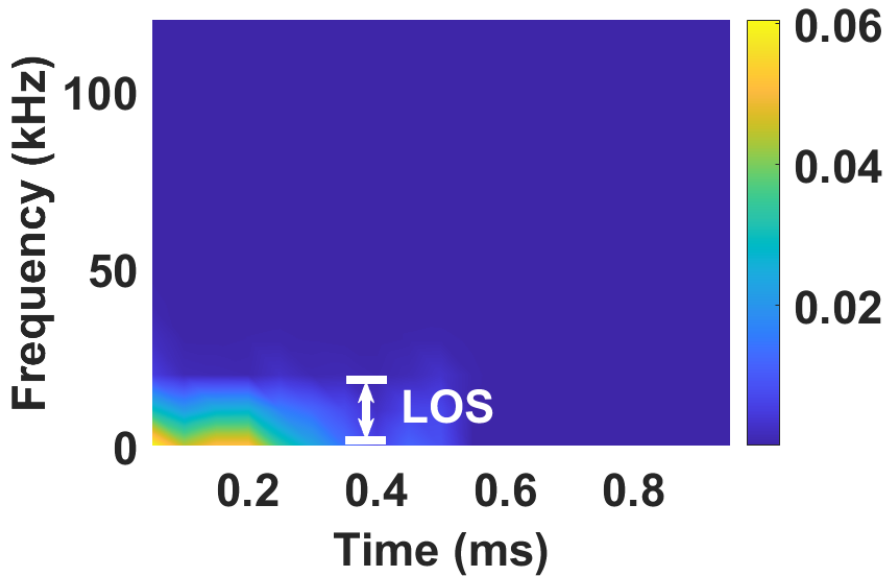}\vspace{-3mm}
    \caption{Filtered Signals}
    \label{fig:filtered_multipath}
\end{subfigure}\vspace{-3mm}
\caption{Multipath Suppression Process in STFT. \textnormal{(a) shows the origin chirp. (b) shows the received chirp with multipath interfere. (c) shows the signal after multiplication. (d) shows the signal after filter}}\vspace{-5mm}
\label{fig:unit_frequency_response}
\end{figure}

Fig.~\ref{fig:unit_frequency_response} shows the process of multipath suppression algorithm on the time-frequency domain.

We discuss how to choose the key signal and algorithm parameters: (i) chirp duration $T$, (ii) bandwidth $B$, (iii) start frequency $f_0$, and (iv) low-pass cutoff frequency $f_{\mathrm{cut}}$.

For the duration $T$, shorter chirps are preferred because they reduce temporal overlap between LOS and NLOS components. However, decreasing $T$ also reduces the effective range. The $T$ is determined by experiments. For example, with the hardware in Sec.~\ref{sec:implementation}, we require $T \geq 0.2\,\mathrm{ms}$ to achieve reliable detection beyond $10\,\mathrm{m}$.

For the bandwidth $B$ and start frequency $f_0$, the chirp sweeps from $f_0$ to $f_0 + B$, with center frequency $f_0 + \tfrac{B}{2}$. The usable $B$ is determined by the frequency responses of the signal generator, power amplifier, and PZT; both $f_0$ and $f_0 + B$ must lie within their effective passband. In our prototype, we use $B = 125\,\text{kHz}$ and $f_0 = 125\,\text{kHz}$, giving a center frequency of $187.5\,\text{kHz}$, close to the PZT resonance at $183\,\text{kHz}$ and within its effective operating band.

For the cutoff frequency $f_{\mathrm{cut}}$, we first determine $T$, $B$, and the minimum TDoA $t_{\min}$, estimated from simulations of the target environment. After multiplied as shown in Fig.~\ref{fig:multipled_multipath}, the LOS component is located near baseband, while the earliest NLOS component is centered at frequency offset $k t_{\min}$, where $k = B/T$ is the chirp rate. To preserve the LOS envelope while attenuating NLOS components, we choose
\begin{equation}
  f_{\mathrm{cut}} < k t_{\min},
\end{equation}
so that multipath delays larger than $f_{\mathrm{cut}}/k$ are effectively suppressed. Increasing the chirp rate $k$ therefore improves the ability to resolve smaller TDoAs.
For example. if we set the parameters for the scene as depicted~\ref{fig:tdoa_distribution}, we set the following paramters.
\[
T = 0.2\ \mathrm{ms}, 
\quad
B = 125\ \mathrm{kHz}, 
\quad
f_0 = 125\ \mathrm{kHz}, 
\quad
f_{\mathrm{cut}} = 35\ \mathrm{kHz},
\]
which yields a TDoA resolution of $f_{\mathrm{cut}}/k \approx 0.056\ \mathrm{ms}$, is less than the minimal LOS-NLOS TDoA of 0.65~ms.

Overall, both the environment and the hardware must be considered when choosing these parameters.

\subsection{Depth Estimation}
\begin{figure}
    \centering
    \begin{subfigure}[t]{0.25\textwidth}
    \includegraphics[width=\linewidth]{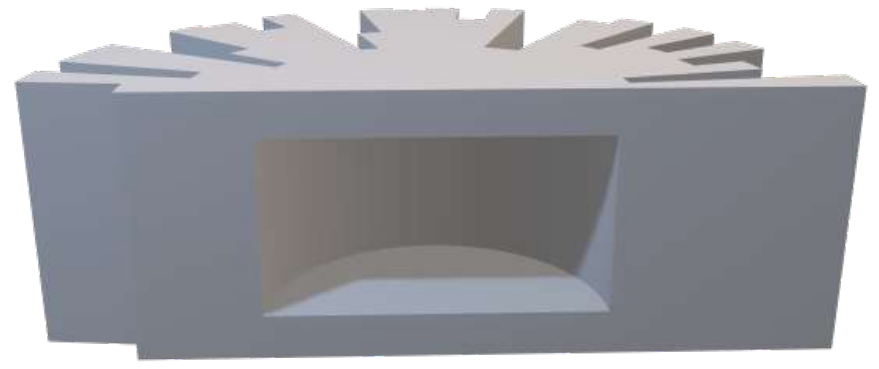}
    \caption{AMs side view}
    \label{fig:depth_side_view}        
    \end{subfigure}
    \begin{subfigure}[t]{0.5\textwidth}
    \includegraphics[width=\linewidth]{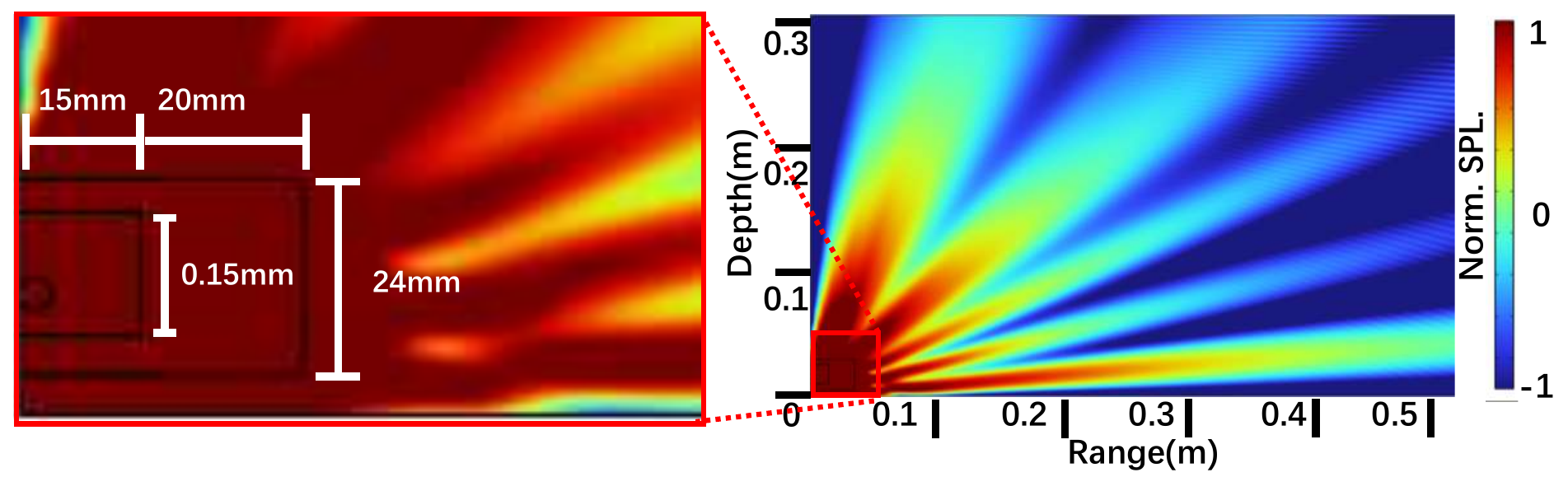}
    \caption{The sound radiation pattern in the vertical plane}\vspace{-3mm}
    \label{fig:depth_spl}        
    \end{subfigure}
    \caption{Vertical Plane Simulation of AMS}\vspace{-3mm}
    \label{depth_simulation}
\end{figure}

To estimate depth, we exploit the elevation-dependent spectral diversity introduced by the AMS. Although the AMS is primarily designed for azimuthal directivity, its structure is not vertically uniform (see Fig.~\ref{fig:depth_side_view}), which naturally induces angle-dependent responses in the vertical plane.
We validate this effect through simulation. As shown in Fig.~\ref{fig:depth_spl}, the sound pressure level (SPL) varies systematically with the elevation angle: different elevation angles exhibit distinct SPL distributions. This elevation-dependent SPL pattern indicates that the received spectrum encodes the elevation angle, which we subsequently use as a cue for depth estimation.

\label{app:depth_estimation}
\subsection{Demo: A 3D moving localization}
\label{app:demo}
\begin{figure}[h!]
    \centering    \includegraphics[width=0.25\textwidth]{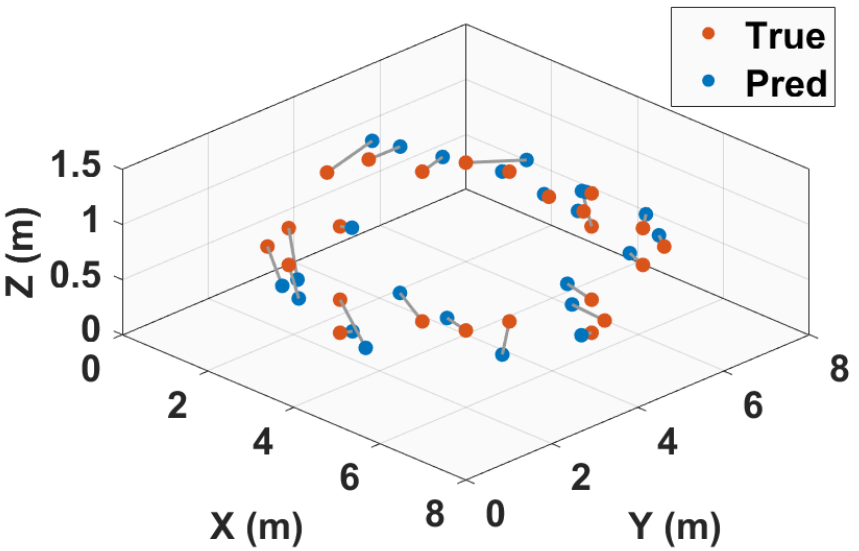}
    \caption{Demo Trajectory}
\label{fig:demo_trajectory}
\end{figure}

We further present a moving trajectory demo. The receiver is manually moved at an approximate speed of 0.2 m/s to form a continuous path. Using measurements from a single anchor, we estimate the receiver’s trajectory and compare it against the ground-truth path. As shown in Fig.~\ref{fig:demo_trajectory}, the predicted trajectory closely tracks the actual motion, demonstrating that our method can achieve accurate localization for moving receivers even with only one anchor.

\end{document}